\documentclass[10pt, conference]{IEEEtran}
\IEEEoverridecommandlockouts
\usepackage{cite}
\usepackage{amsmath,amssymb,amsfonts}
\usepackage{algorithmic}
\usepackage{graphicx}
\usepackage{textcomp}
\usepackage{xcolor}

\usepackage{multirow}
\usepackage{booktabs}
\usepackage{subcaption}
\usepackage{comment}
\usepackage{balance}
\usepackage{enumitem}
\usepackage{algorithm}
\usepackage{colortbl}
\definecolor{mygray}{gray}{0.9}

\graphicspath{{figures/}}

\def\BibTeX{{\rm B\kern-.05em{\sc i\kern-.025em b}\kern-.08em
    T\kern-.1667em\lower.7ex\hbox{E}\kern-.125emX}}

\begin{document}
\title{GesturePrint: Enabling User Identification for mmWave-based Gesture Recognition Systems}

\author{%
    \IEEEauthorblockN{Lilin Xu\IEEEauthorrefmark{1}, Keyi Wang\IEEEauthorrefmark{2}, Chaojie Gu\IEEEauthorrefmark{1}, Xiuzhen Guo\IEEEauthorrefmark{1}, Shibo He\IEEEauthorrefmark{1}, Jiming Chen\IEEEauthorrefmark{1}}
    \IEEEauthorblockA{
        \textit{\IEEEauthorrefmark{1}Zhejiang University, \IEEEauthorrefmark{2}University of California San Diego} 
    }
    \IEEEauthorblockA{
        lilinxu@zju.edu.cn, kew035@ucsd.edu, \{gucj, guoxz, s18he, cjm\}@zju.edu.cn
    }
}

\maketitle
\begin{abstract}
The millimeter-wave~(mmWave) radar has been exploited for gesture recognition. 
However, existing mmWave-based gesture recognition methods cannot identify different users, which is important for ubiquitous gesture interaction in many applications.
In this paper, we propose \textit{GesturePrint}, which is the first to achieve gesture recognition and gesture-based user identification using a commodity mmWave radar sensor. 
\textit{GesturePrint} features an effective pipeline that enables the gesture recognition system to identify users at a minor additional cost.
By introducing an efficient signal preprocessing stage and a network architecture GesIDNet, which employs an attention-based multilevel feature fusion mechanism, \textit{GesturePrint} effectively extracts unique gesture features for gesture recognition and personalized
motion pattern features for user identification.
We implement \textit{GesturePrint} and collect data from 17 participants performing 15 gestures in a meeting room and an office, respectively. 
\textit{GesturePrint} achieves a gesture recognition accuracy~(GRA) of 98.87\% with a user identification accuracy~(UIA) of 99.78\% in the meeting room, and 98.22\% GRA with 99.26\% UIA in the office.
Extensive experiments on three public datasets and a new gesture dataset show \textit{GesturePrint}'s superior performance in enabling effective user identification for gesture recognition systems.
\end{abstract}

\maketitle

\section{Introduction}

\label{sec:intro}
Nowadays, the mmWave radar has received increasing attention from industry and academia because of its low power, high spatial resolution, and robustness to temperature and lighting conditions. The mmWave radar has empowered plenty of applications in autonomous driving~\cite{chadwick2019distant,chang2020spatial}, human localization and tracking~\cite{zhao2019mid,wu2020mmtrack}, activity recognition~\cite{singh2019radhar,xu2023mesen} and health care~\cite{yang2016monitoring,chen2021movi,gong2021rf}.
Recently, there has been a trend of utilizing mmWave radars to implement gesture recognition systems, which enables ubiquitous gesture interaction in a broad spectrum of applications, including gaming control, Internet of Things (IoT), and virtual reality (VR).
Compared with traditional methods~\cite{laput2016viband,chen2019taprint} based on wearable devices, mmWave radars enable touchless gesture recognition without causing discomfort or extra burden to users. In addition, mmWave-based approaches are more privacy-preserving than visual-based solutions~\cite{guo2022mudra}.
As for other RF-based solutions like using WiFi~\cite{abdelnasser2015wigest,virmani2017position}, mmWave sensing operates at extremely high frequency (EHF) bands, and it has a higher spatial resolution for fine-grained motion detection. 

\begin{figure}[t]
    \centering
    \begin{subfigure}[b]{0.48\textwidth}
        \centering
        \includegraphics[width=1.\linewidth]{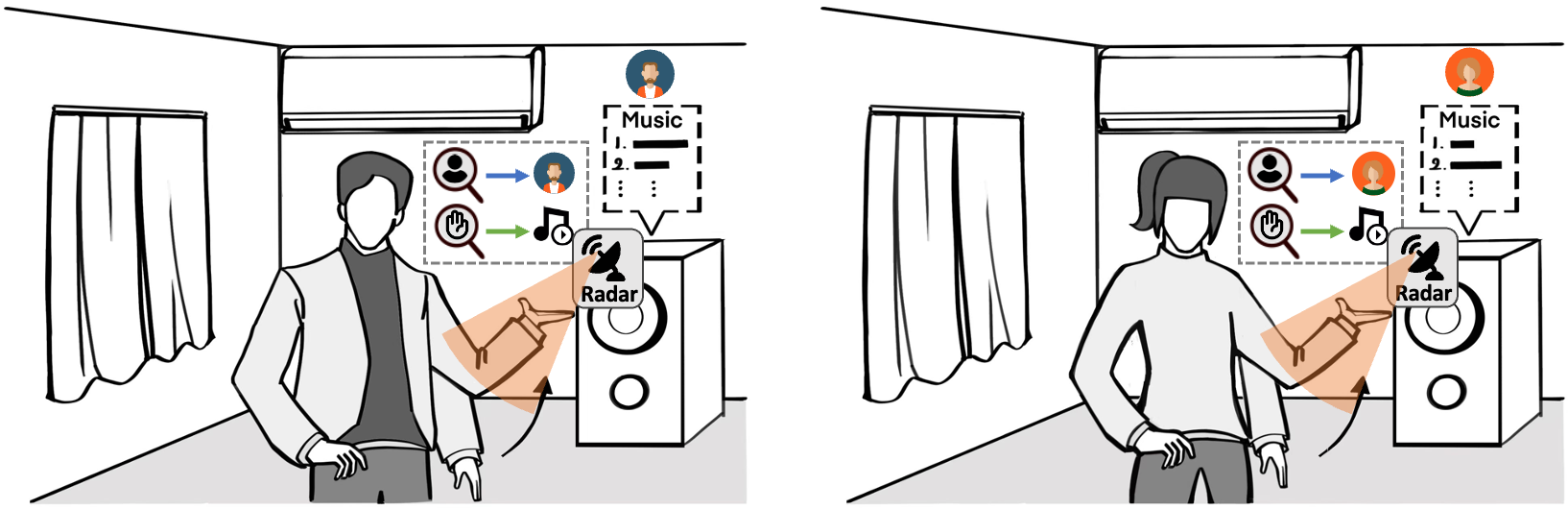}
        \caption{Controlling the music player to list songs based on one's taste.}
        \label{fig:application_1}
    \end{subfigure}
    \hfill
    \begin{subfigure}[b]{0.48\textwidth}
        \centering
        \includegraphics[width=1.\linewidth]{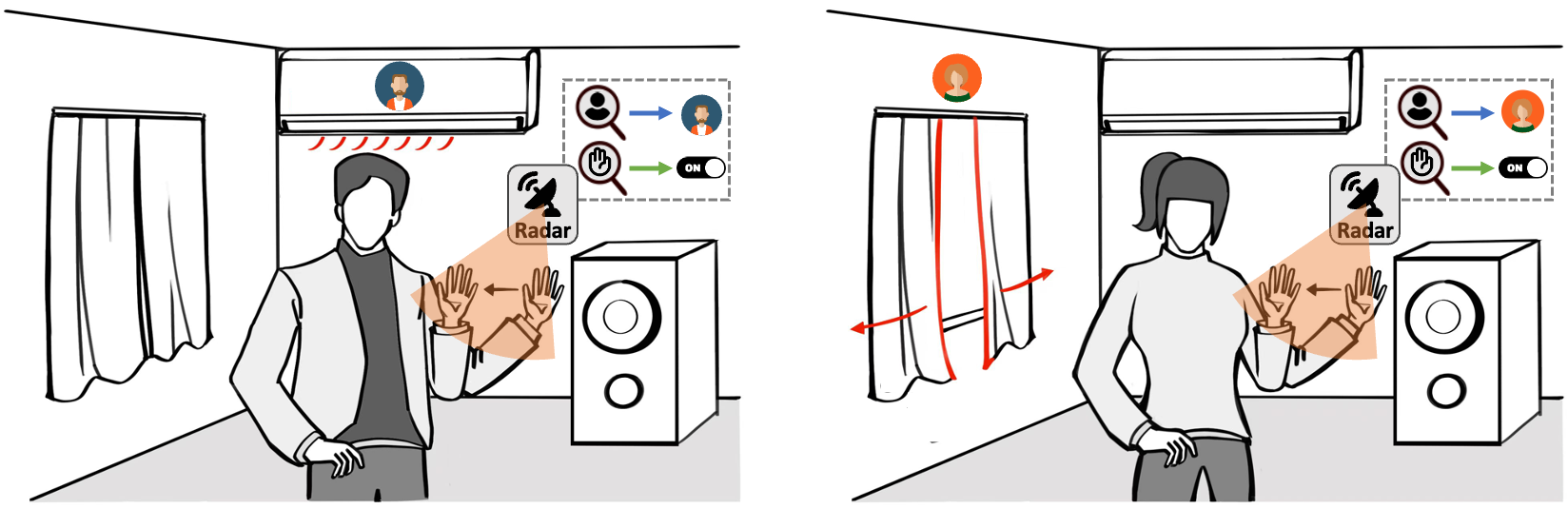}
        \caption{Personalizing the meaning of gestures to operate home devices.}
        \label{fig:application_2}
    \end{subfigure}
     \caption{Potential applications of mmWave-based gesture recognition systems with user identification capability. The identification capability can significantly improve the user experience in interacting with smart devices.}
    \label{fig:application}
\end{figure}

In the last several years, several mmWave-based gesture recognition solutions~\cite{liu2020real,liu2022mtranssee,palipana2021pantomime,salami2022tesla} have been proposed.
Although existing solutions can accurately recognize predefined gestures performed by different users, none of them can identify the user who performs the gestures. In practice, the capability of user identification can significantly improve the user experience in interacting with smart devices. For example, as shown in Fig. \ref{fig:application_1}, when a user performs a predefined gesture to turn on a music player using a mmWave-based controller, the controller obtains the user's identity and instructs the player to list songs according to the user's taste. Also, the user can personalize the meaning of gestures, e.g., waving one hand from left to right to open/close the curtain or decrease/increase the air conditioning temperature as demonstrated in Fig. \ref{fig:application_2}.

While there exist some gesture-based user identification solutions that utilize other modalities such as video~\cite{guo2022mudra}, IMU~\cite{yu2020thumbup}, and WiFi~\cite{li2020wihf}, they cannot be directly applied to mmWave sensing due to the differences in data format and feature representation across modalities. Additionally, the mmWave radar has unique application scenarios that differ from other modalities. Thus, solutions specially designed for mmWave-based point clouds are essential to fully exploit the potential of gesture-based interaction with the mmWave radar.

To unleash the potential of mmWave-based gesture recognition systems, we propose \textit{GesturePrint}, an effective one-stop solution to enable the gesture recognition systems to identify users at a minor additional cost. \textit{GesturePrint} first obtains point clouds related to objects in the environment by the radar device.
It then segments the gesture motions utilizing a parameter-adaptive sliding window method, subsequently removing the outlier points that are distinguished from the target points reflected from people. With the preprocessed data, \textit{GesturePrint} can accurately recognize predefined gestures and identify the user who performs the gestures based on our specially designed network architecture GesIDNet.

To achieve this, there are several practical challenges in designing \textit{GesturePrint}, as summarized below.

\noindent\textbf{Effective Gesture Capture}: Reliable gesture segmentation is essential, as it influences the amount of relevant information and noise preserved for subsequent gesture recognition and user identification. However, determining the beginning and end of a gesture is challenging due to the sparsity and irregularity of point clouds obtained by the radar. To address this issue, \textit{GesturePrint} employs a parameter-adaptive sliding window method to effectively detect the start and the end of a gesture motion. 
Furthermore, in real-world scenarios, there are many noise points from the surroundings. Even in the same environment, the distribution of noise points changes considerably when the surrounding objects move slightly. This adversely affects the processing of gesture-related points and consequently degrades the system's performance on gesture recognition and user identification. 
To mitigate this issue, \textit{GesturePrint} eliminates outlier points unrelated to the user throughout the entire gesture motion as much as possible.

\noindent\textbf{Effective Feature Extraction}: On the one hand, achieving a one-stop solution for recognition and identification requires that the method can obtain effective information from the point cloud data to distinguish distinct gestures and different users, respectively. 
On the other hand, it is challenging to extract effective and reliable gesture features and user-personalized motion pattern features from the sparse point clouds captured by the mmWave radar sensor. To solve the problems above, we introduce a network architecture, GesIDNet, which is specially designed to accommodate the characteristics of gesture point clouds. GesIDNet features an attention-based multilevel feature fusion module that adaptively fuses low-level features and high-level features extracted from the sparse point clouds to obtain effective features.

\noindent\textbf{Robust Feature Learning}: 
There are inevitable differences among multiple repetitions of the same gesture performed by the same user, e.g., different motion speeds and different distances from the mmWave radar sensor. It is challenging for the system to stay robust to these differences.
The design of GesIDNet contributes to mitigating the impact brought by this issue.
Besides, \textit{GesturePrint} employs data augmentation during the training phase to enhance the system's robustness.

In \textit{GesturePrint}, we utilize a commodity mmWave radar sensor and a laptop for data collection and inference. We use a back-end server to train the models. We design a signal processing pipeline to segment gestures from the collected data. GesIDNet then extracts effective features from the segmented gesture data for gesture recognition and user identification. We conduct extensive experiments on three public datasets and a new gesture dataset to evaluate \textit{GesturePrint} and compare its performance with other state-of-the-art methods. 
While existing mmWave-based point cloud gesture datasets only contain self-defined gestures, we build a new dataset of standard gestures in ASL~(American Sign Language). 
Experimental results show that \textit{GesturePrint} achieves an overall performance of 98.87\% accuracy for gesture recognition and 99.78\% accuracy for gesture-based user identification with 15 gestures and 17 users in the meeting room environment while 99.88\% for recognition and 97.60\% for identification with 5 gestures and 32 users. 
Besides, it achieves an average result of 0.75\% EER~(Equal Error Rate) for user identification across all datasets.
We further evaluate \textit{GesturePrint} under different experimental settings, and all the results indicate that it is effective for both gesture recognition and user identification.
In summary, our major contributions are summarized as follows:
\begin{itemize}[leftmargin=*]
\item To the best of our knowledge, \textit{GesturePrint} is the first solution that augments mmWave-based gesture recognition systems with user identification capability. 
\item We introduce an efficient signal preprocessing stage and a specially designed network architecture GesIDNet to achieve effective feature extraction from sparse point clouds captured by the radar.
The features include unique gesture features for gesture recognition and personalized motion pattern features for user identification.
\item We build a mmWave-based gesture dataset, including 9,332 samples from 17 participants performing 15 ASL gestures in two different environments. Besides, we conduct extensive experiments to evaluate \textit{GesturePrint} on three public gesture recognition datasets.
The results demonstrate \textit{GesturePrint}'s superior performance in enabling effective user identification for gesture recognition systems.
\end{itemize}

The rest of this paper is organized as follows.
\S\ref{sec:related_work} reviews related studies.
\S\ref{sec:feasibility_analysis} presents the preliminary study.
\S\ref{sec:design} introduces the detailed design of \textit{GesturePrint}. \S\ref{sec:implementation} and \S\ref{sec:evaluation} present the implementation and evaluation results.
\S\ref{sec:discussion} discusses some related issues.
\S\ref{sec:conclusion} concludes this work.
\section{Related work}
\label{sec:related_work}
\subsection{mmWave-based Gesture Recognition} 
mmWave-based gesture recognition demonstrates great potential in interaction applications~\cite{liu2020real,liu2022mtranssee,palipana2021pantomime,salami2022tesla}, as mmWave radars provide fine-grained resolution and easy deployment.
Due to the collected data with high-dimensional features, existing methods mainly employ neural networks for effective feature extraction.
According to the input data format, existing approaches can be broadly divided into \textit{signal map-based} and \textit{point cloud-based}.

\noindent\textbf{Signal map-based Approaches:} mmASL~\cite{santhalingam2020mmasl} proposes a multi-task deep learning architecture to extract gesture domain features from the spectrograms.
DI-Gesture~\cite{li2021towards} uses a dynamic window mechanism and a network combined by CNN and LSTM to achieve domain-independent and real-time gesture recognition. 
RadarNet~\cite{hayashi2021radarnet} proposes a novel CNN architecture dealing with range-Doppler maps to recognize swipe gestures.
M-gesture~\cite{liu2021m} designs a compact CNN architecture to suppress the influence of different users, and achieves person-independent finger gesture recognition.

\noindent\textbf{Point cloud-based Approaches:} Due to advancements in handling unordered point clouds, point cloud-based methods have recently become a research trend in mmWave-based gesture recognition.
mHomeGes~\cite{liu2020real} and mTransSee~\cite{liu2022mtranssee} both convert point clouds into the concentrated position-doppler profile to emphasize the positional relationship and speed differences among points, and use designed convolutional networks to extract gesture features.
Pantomime~\cite{palipana2021pantomime} combines PointNet++ and LSTM to extract spatio-temporal features from mmWave radar point clouds, while Tesla-Rapture~\cite{salami2022tesla} utilizes a temporal K-NN module based on graph convolution.

Different from existing studies, apart from extracting unique gesture features for gesture recognition, \textit{GesturePrint} can effectively capture personalized motion pattern features from the gesture point clouds by employing an efficient signal preprocessing stage and a specially designed network architecture. In this way, it empowers mmWave-based gesture recognition systems with the capability of user identification.

\subsection{mmWave-based User Identification} 
As mmWave radars can capture human biometric information with fine-grained spatial resolution, some mmWave-based user identification methods have been proposed in recent years.

VocalPrint~\cite{li2020vocalprint} uses a mmWave radar sensor to capture skin-reflect signals around the near-throat region during speech, while the work in~\cite{dong2020secure} utilizes both vocal cord vibration and lip motion. 
Moreover, M-Auth~\cite{wang2022your} achieves multi-user authentication by capturing unique breathing patterns from radar signals.
Additionally, researchers have explored gait-based identification.
MU-ID~\cite{yang2020mu} identifies multiple users simultaneously by capturing gait biometrics. mID~\cite{zhao2019mid} uses 3-D voxel grids and a deep recurrent network for identification and tracking. mmGait~\cite{meng2020gait} proposes a CNN network directly using point clouds as input. SRPNet~\cite{cheng2021person} employs a network that combines PointNet and Bi-LSTM to extract spatio-temporal information from point clouds.

However, these mmWave-based identification methods cannot be directly applied for user identification with gestures due to the heterogeneity in different motion patterns.
Integrating these methods into existing gesture recognition systems to achieve user identification will add an extra motion burden to the user.
To the best of our knowledge, \textit{GesturePrint} is the first work that explores the feasibility of identifying users based on their gesture motions to further unleash the potential of gesture-based interaction with the mmWave radar at a minor extra cost.
This one-stop solution for recognizing gestures and identifying users can enhance the applications of mmWave-based gesture interactions.
\begin{figure}[t]
    \centering
    \includegraphics[width=0.48\textwidth]{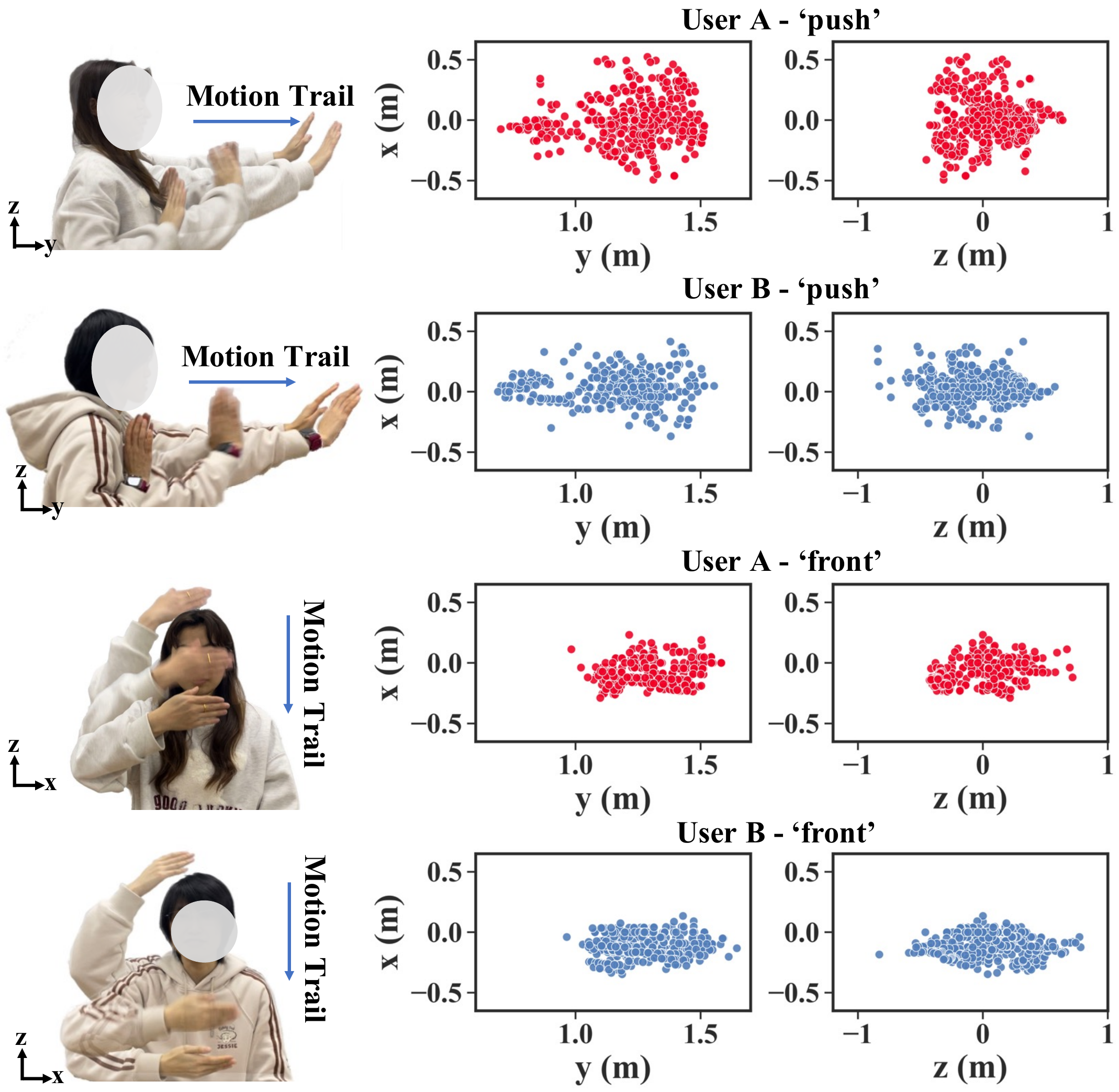}
    \caption{The visualization of gesture point clouds obtained from User A and User B when they perform ASL gestures.
    }
    \label{fig:feasibility_gestures}
\end{figure}

\section{Preliminaries}
\label{sec:feasibility_analysis}
The mmWave radar captures signals reflective of gesture motions as users perform gestures. 
As personal behavioral traits can serve as biometrics~\cite{pfeuffer2019behavioural}, we utilize a mmWave radar sensor to capture such characteristics, including behavior manners and personalized unconscious motion styles, from gesture motions for user identification.

mmWave point clouds are generated through a series of signal processing steps applied to the frequency modulated continuous wave (FMCW) signal reflected by objects, which contains information about the range, velocity, and angle of the objects. 
The processing steps include Range Fast-Fourier Transform~(FFT), Doppler FFT, Constant False Alarm Rate (CFAR), and Angle FFT.
The resulting point clouds contain crucial information about the objects, which offers insights into their shapes and movements.
Fig.~\ref{fig:feasibility_gestures} shows the visualization of point clouds captured from User A and User B when they performed ASL signs `push' and `front'. 
These two users have similar body shapes, with a height of around $160\,\text{cm}$ and a weight of around $48\,\text{kg}$.
The gesture point clouds exhibit characteristics that can be utilized for gesture recognition and user identification, respectively.
On the one hand, point clouds can demonstrate different shapes and movements of different ASL sign gestures. 
On the other hand, it is observed that gesture point clouds can differ in space and time between the same gesture performed by different users, such as point number, coverage, and density. These differences are mainly caused by individual variations in arm length, motion speed, range of motion, and even implicit motion habits.
For instance, as shown in Fig.~\ref{fig:feasibility_gestures}, the range of User B's point clouds on the x-axis is narrower than that of User A, implying that User B has a more limited range of motion on the x-axis when executing gestures.
However, compared with the differences in point clouds resulting from distinct gestures, the differences between point clouds of the same gesture performed by different users are not as pronounced, indicating that gesture-based user identification is a more challenging task.

\begin{figure}[t]
    \centering
    \includegraphics[width=0.485\textwidth]{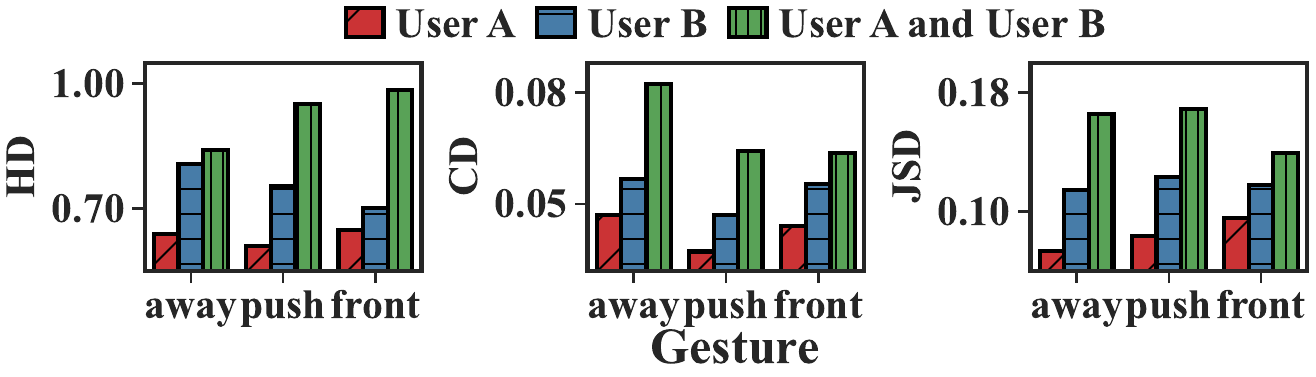}
    \caption{Differences among point clouds are measured by three metrics: Hausdorff distance (HD), Chamfer distance (CD), and Jensen-Shannon divergence (JSD).}
    \label{fig:similarity_dis}
\end{figure}
\begin{figure*}[t]
    \centering
    \includegraphics[width=0.99\textwidth]{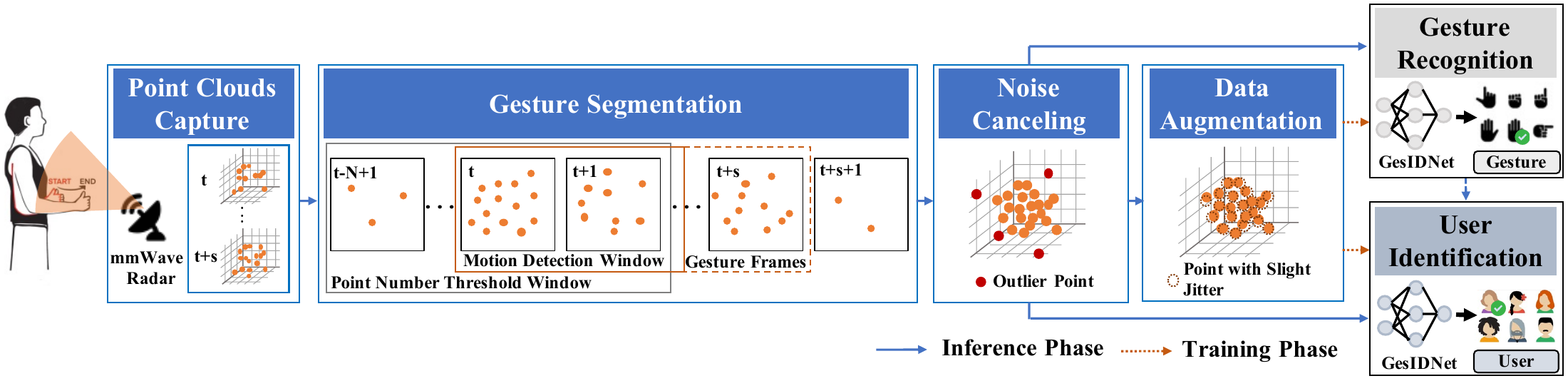}
    \caption{Overview of \textit{GesturePrint}: a system that enables mmWave-based gesture recognition and gesture-based user identification, comprising the data preprocessing and classification stages.
    }
    \label{fig:overview}
\end{figure*}

To further study the feasibility of identifying users based on gesture point cloud data captured by the radar, we utilize three metrics, i.e., Hausdorff distance (HD)~\cite{berger2013benchmark}, Chamfer distance (CD)~\cite{fan2017point}, and Jensen-Shannon divergence (JSD)~\cite{achlioptas2018learning}, to evaluate gesture points clouds.
The HD measures the extent to which each point in one point cloud lies near some point in another point cloud;
the CD measures the average bidirectional closest-point distance between two distinct point clouds;
the JSD measures the degree to which points in one point cloud tend to occupy similar locations as those of another point cloud.
Let $\boldsymbol{G} = \{g_1, g_2, ..., g_{N_g}\}$ denote $N_g$ different gestures.
For each user, there are $N_c$ point clouds $\boldsymbol{C}^{g_i} = \{c_{1}^{g_i}, c_{2}^{g_i}, ...,c_{N_c}^{g_i}\}$ for gesture $g_i \in \boldsymbol{G}$ after the preprocessing stage (see \S\ref{subsec:dataprocessing} for more details).
Then, we measure the difference between point clouds as
\begin{equation}
    d(g_i) = \frac{{\sum_{m=1}^{N_{c2}}\sum_{n=1}^{N_{c1}}}D({c}_n^{g_i},{c}_m^{g_i})}{{N_{c1}}{N_{c2}}},
    \label{eq:pc_distance}
\end{equation}
where ${c}_n^{g_i} \in \boldsymbol{{C}}^{g_i}_{1}$, ${c}_m^{g_i} \in \boldsymbol{{C}}^{g_i}_{2}$ and ${c}_n^{g_i} \neq {c}_m^{g_i}$. 
$D(\cdot)$ denotes the metric utilized for measurement, and it can be either $HD(\cdot)$ or $CD(\cdot)$ or $JSD(\cdot)$.
When measuring the difference between the same gesture performed by the same user, $\boldsymbol{{C}}^{g_i}_{1}$ and $\boldsymbol{{C}}^{g_i}_{2}$ in Eq.~\ref{eq:pc_distance} are from the same user; when measuring the differences between the same gesture performed by different users, $\boldsymbol{{C}}^{g_i}_{1}$ and $\boldsymbol{{C}}^{g_i}_{2}$ are from different users.

For a fair comparison, both User A and User B are instructed to execute 3 ASL gestures~(`away', `push', and `front'), with each gesture being performed 10 times.
The HD, CD, and JSD results of the gesture point clouds are displayed in Fig.~\ref{fig:similarity_dis}. 
These results demonstrate that, for the same ASL gesture, the differences between gesture point clouds from different users are more noticeable than those from the same user.
This suggests that gesture point clouds captured by the radar contain information that can be used to differentiate between users when they perform gesture motions.
Although the mmWave radar can sense these differences, it is difficult to directly distinguish distinct gestures and different users from the point clouds since the point clouds are abstract and contain high-dimensional features, which cannot be fully exploited through only calculating distances and divergence between the point clouds.
Thus, it is desirable to design an efficient data preprocessing method and an effective network architecture, which can extract unique gesture features and features containing personalized motion patterns from point clouds, to effectively achieve both gesture recognition and user identification.
\section{GesturePrint Design}
\label{sec:design}
In this section, we first provide an overview of \textit{GesturePrint}.
Then, we present the detailed design of \textit{GesturePrint}.

\subsection{\textbf{System Overview}}
\label{subsec:system}
\textit{GesturePrint} is designed to extract effective features regarding specific gestures and users from sparse point clouds captured by the mmWave radar. Fig.~\ref{fig:overview} demonstrates the system pipeline of \textit{GesturePrint}, which has two major stages with six modules. 
The data preprocessing stage includes \textit{point clouds capture}, \textit{gesture segmentation}, \textit{noise canceling}, and \textit{data augmentation}. The classification stage includes \textit{gesture recognition} and \textit{user identification}. \textit{GesturePrint} works with a commodity mmWave radar sensor, after obtaining the points converted from signal data through the radar, it segments out gestures from temporal point cloud frames by using an adaptive sliding window method. 
After gesture segmentation, \textit{GesturePrint} discards outlier noise points that are not reflected from the human body. 
After noise canceling, the points captured by the radar in the whole gesture process are aggregated as the gesture point cloud, which is then fed into GesIDNet, a specially designed network architecture, for gesture recognition.
With the recognition result, GesIDNet is employed to further identify the user performing the gesture with the gesture-corresponding recognition model. Finally, the gesture and the user are both inferred by \textit{GesturePrint}. In particular, during training, we augment the data by adding some random jitters to the points. Although the models for the two tasks are trained individually, and the features extracted for the two tasks are different, \textit{GesturePrint} does not need any extra data compared with the stand-alone gesture recognition task. We use the same data to dig for more information from another dimension for user identification.

\subsection{\textbf{Data Preprocessing Stage}}
\label{subsec:dataprocessing}
In this stage, \textit{GesturePrint} processes the raw data captured by the radar to obtain gesture point clouds, which are then utilized as the input in the subsequent classification stage.

\noindent\textbf{Point Clouds Capture}: The transformation from raw signal data reflected by objects in the environment into point cloud data format is done by the mmWave radar device. 
Besides, to mitigate the influence of reflections from surroundings in the environment that are unrelated to the user's activities, we enable the static clutter removal function before object detection. In this way, the objects detected at the zero Doppler velocity bins by the device can be discarded.

\noindent\textbf{Gesture Segmentation}: This module segments complete gesture motions from the captured point cloud data. 
Unlike DI-Gesture~\cite{li2021towards} segmenting gestures by applying a dynamic window mechanism to DRAI~(Dynamic Range Angle Image) captured by the radar, we segment gestures based on radar point clouds.
We exploit the number of points in each frame and design a parameter-adaptive sliding window method to detect which frame a gesture motion starts and ends.
\textit{GesturePrint} first calculates the cumulative distribution density of points over $N$ frames and obtains a dynamic point number threshold $P_{Thr}$ based on the distribution. It then uses a sliding motion detection window of length $n$ with $P_{Thr}$ to determine whether the current frame contains gesture motion or not. Once a frame is determined as a motion frame, \textit{GesturePrint} checks the motion frame count in the sliding window. If the number exceeds the minimum frame threshold $F_{Thr}$, the frame is regarded as the start of the motion, and the subsequent frames are added to the motion frame group until all frames in the sliding window are determined as static frames, signaling the end of the gesture motion.
The points in the motion frame group are then aggregated.

\noindent\textbf{Noise Canceling}: Although some noise reflected from the environment can be removed by enabling the static clutter removal function during data collection, the remaining points are not all from the user's gesture motion.
There are still some noise points caused by reflectors' subtle movement or signal multipath reflection.
These noise points indeed adversely affect gesture recognition and user identification. To better obtain user-related points and mitigate the influence of noise, \textit{GesturePrint} apply DBScan, a density-aware algorithm, to cluster points based on pairwise distances within the whole point cloud.
We set the parameters for DBScan to get point clusters, including the maximum distance between pair points $D_{max}$, and the minimum point number of a point cluster $N_{min}$.
Among all the clusters obtained through DBScan, the cluster containing most of the points is retained as the main cluster, while others are discarded. 
This main cluster, related to the human body and regarded as the gesture point cloud, is then prepared for being fed into GesIDNet.

\noindent\textbf{Data Augmentation}: 
To enhance the system's robustness, data augmentation is employed during the training phase.
We introduce a random subtle displacement $j$ to each point $p$ in the gesture point cloud $P$.
This process is repeated to augment the data three times.
Displacements for each gesture point cloud are generated using a Gaussian distribution with the mean $\mu=0$ and the standard deviation $\sigma=0.02$.

\begin{figure*}[t]
    \centering
    \includegraphics[width=1.\textwidth]{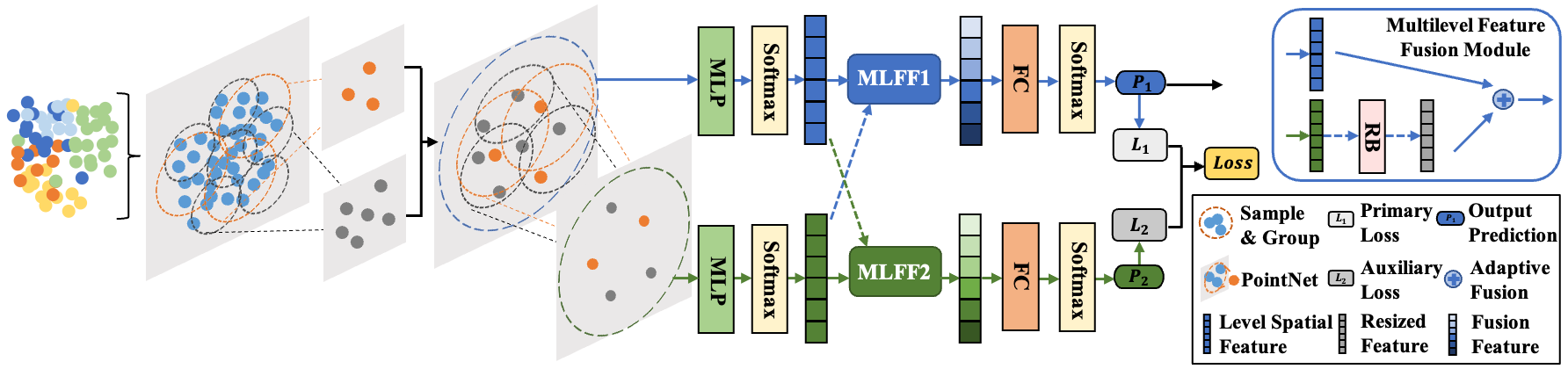}
    \caption{The architecture of GesIDNet. MLP: Multilayer Perceptron, FC: Fully Connected Layer, RB: Resizing Block.}
    \label{fig:GesIDNet}
\end{figure*}
\subsection{\textbf{Classification Stage}}
In this stage, we propose GesIDNet~(as demonstrated in Fig.~\ref{fig:GesIDNet}) to address recognition and identification tasks based on the same gesture point cloud obtained in the data preprocessing stage.
For the gesture recognition task, GesIDNet can effectively extract unique features for different gestures.
For the user identification task, GesIDNet can extract the high-dimensional features related to personalized motion patterns, which represent the motion styles of different users. The same data are fed to GesIDNet with different labels, i.e., gesture labels and user identity labels, to train gesture recognition models and user identification models, respectively. 

At runtime, there are two identification modes in our system, i.e., \textit{serialized mode} and \textit{parallel mode}. The serialized mode identifies users with the result of gesture recognition. The gesture point cloud obtained in the preprocessing stage is fed into the GesIDNet recognition model to recognize the gesture. 
Based on the inferred gesture, \textit{GesturePrint} selects the corresponding identification model for further user identification.
In contrast, the parallel mode conducts gesture recognition and user identification separately. The main difference between these two modes is that during training time, the user identification model is trained for every gesture separately or trained with all predefined gestures. According to the overall performance of these two modes~(see \S\ref{sec:evaluation} for more details) and the capability of handling random gestures and unauthorized people, we set the serialized mode as our default identification mode.
Finally, with the outputs of GesIDNet, \textit{GesturePrint} achieves gesture recognition and user identification. The design details of GesIDNet are described below.

Fig.~\ref{fig:GesIDNet} demonstrates the design of GesIDNet. 
PointNet++~\cite{qi2017pointnet++} can extract details and features from data structured in point cloud format.
However, while PointNet++ is typically employed with large-scale dense point clouds, the gesture point clouds obtained by the mmWave radar are usually sparse. Thus, to effectively extract features from the sparse gesture point clouds, we adopt the set abstraction block of PointNet++ and further design a multilevel feature fusion module with an attention mechanism.

On the one hand, since the sparse points in each frame captured by the radar provide insufficient information for fine-grained local features extraction~\cite{palipana2021pantomime}, we aggregate points captured by the radar in the whole gesture process following the gesture segmentation module, and the aggregated points are further processed as the input of GesIDNet.
Although we aggregate all the frames together, owing to noise canceling, point clouds in every frame are usually in a relatively concentrated place in the space. Therefore, by extracting and combining local features, a part of the information in the frame dimension is still preserved, which facilitates the effective extraction of gesture features and user-personalized motion pattern features.
GesIDNet employs the set abstraction block of PointNet++ to extract local spatial features at different scales from the aggregated gesture point clouds, and these multiscale local features are combined for extracting higher-level features. To be specific, one single set abstraction block can be described as follows: with a set of unordered points $\{p_1, p_2, ..., p_N\}$ as the input, GesIDNet samples $n_i$ points from the point set, and groups the nearest $m_i$ points within the radius $d_i$ around them as representation points $\{p^{i}_{1}, p^{i}_{2}, ..., p^{i}_{n_i}\}$ of local regions, then it uses multilayer perceptron (MLP) to extract features. $n_i$, $m_i$, and $d_i$ are hyperparameters of the set abstraction block $i$, and local features $\boldsymbol{f}_{i}$ of different scales are combined as $\boldsymbol{f}_{s}$, which is the output of the block.

On the other hand, the aggregated gesture point clouds comprise an unordered set of points with varying numbers and strong spatio-temporal correlation. 
To exploit the data characteristics, we introduce an attention-based multilevel feature fusion module to adaptively combine low-level features and high-level features extracted from the point clouds.
With the module, GesIDNet adaptively assigns large weights to effective features for gesture recognition and user identification.
To be specific, GesIDNet extracts level feature $\boldsymbol{F}$ from $\boldsymbol{f}_{s}$ by grouping all the representation points and applying MLP. There are two feature levels, i.e., $l_1$~(low-level) and $l_2$~(high-level), in GesIDNet.
The sizes of feature vectors in each level are different, so before fusing features of different levels, in each level, GesIDNet resizes the other level's feature vectors to match the feature vector size at the current level by using a resizing block.
The block includes a linear layer followed by a ReLu layer. Let $\boldsymbol{F}^{k}$ denote the combined feature of point set abstraction at level $k$, $\boldsymbol{F}^{l->k}$ denote the feature $\boldsymbol{F}^{l}$ resized from level $l$ to level $k$ to match the size of the combined feature at level $k$. The fusion feature $\boldsymbol{Y}^{k}$ of $\boldsymbol{F}^{l->k}$ and $\boldsymbol{F}^{k}$ at level $k$ can be described as
\vspace{-0.2em}
\begin{equation}
\boldsymbol{Y}^{k}=S(\boldsymbol{F}^{l->k})\cdot \boldsymbol{F}^{l->k} + S(\boldsymbol{F}^{k})\cdot F^{k},
\end{equation}
where $S(\boldsymbol{F}^{l->k})$ and $S(\boldsymbol{F}^{k})$ refer to the adaptive weights of different level features at level $k$. $S(\cdot)$ denotes a softmax function, and $S(\boldsymbol{F}^{l->k})$ and $S(\boldsymbol{F}^{k})$ can be expressed by
\vspace{-0.2em}
\begin{equation}
\begin{split}
S(\boldsymbol{F}^{l->k})=\frac{e^{g(\boldsymbol{F}^{l->k})}}{e^{g(\boldsymbol{F}^{l->k})}+e^{g(\boldsymbol{F}^{k})}}, \\
S(\boldsymbol{F}^{k})=\frac{e^{g(\boldsymbol{F}^{k})}}{e^{g(\boldsymbol{F}^{l->k})}+e^{g(\boldsymbol{F}^{k})}}.
\label{eq:softmax}
\end{split}
\end{equation}
In Eq.~\ref{eq:softmax}, $S(\boldsymbol{F}^{l->k})+S(\boldsymbol{F}^{k})=1$, $S(\boldsymbol{F}^{l->k})$, $S(\boldsymbol{F}^{k}) \in [0,1]$, and $g(\cdot)$ denotes a convolutional layer. 
In this way, the weights of different level spatial features can be adaptively learned. 

To fully exploit the fusion results at each level, we adopt the idea of auxiliary loss~\cite{ballester2016performance}. The fusion feature $\boldsymbol{Y}^{k}$ at level $k$ is then fed to a couple of FC~(fully connected) layers followed by final classifiers, and the number of FC layers depends on the level. In this way, GesIDNet produces two classification prediction results $P_{1}$ and $P_{2}$ at two feature levels $l_1$ and $l_2$, respectively.
We regard $P_{1}$ as the primary prediction result and the cross-entropy loss calculated from it is called primary loss $L_{1}$ while regarding $P_{2}$ as the auxiliary result with a loss called auxiliary loss $L_{2}$.
During training, we add the auxiliary loss to the primary loss as the final loss, while in the inference phase, we just use the primary classification result as the final prediction result of GesIDNet. 

\begin{figure}[t]
    \centering
    \includegraphics[width=0.49\textwidth]{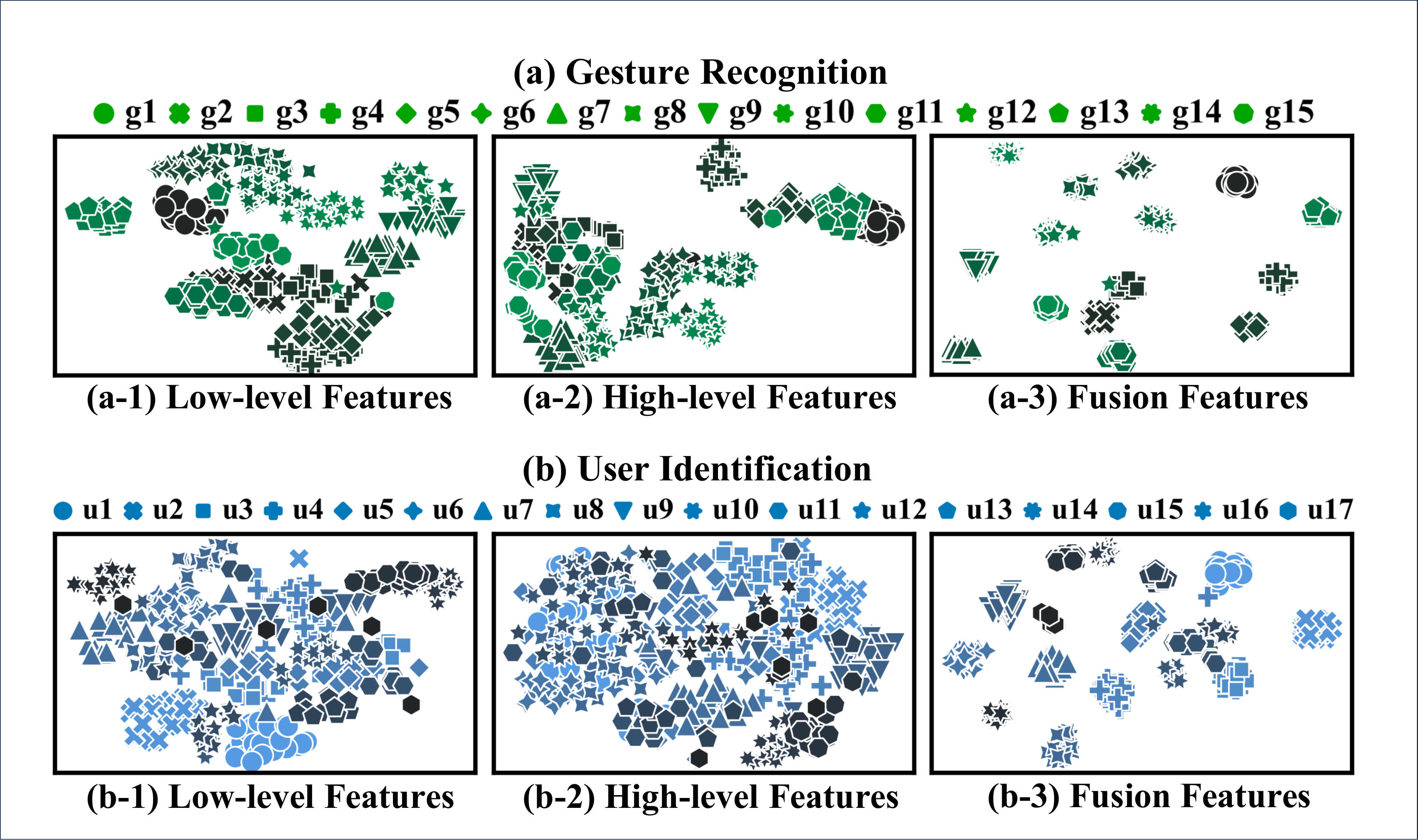}
    \caption{The visualization of features extracted by GesIDNet for gesture recognition and user identification, respectively. Different shapes denote different gestures/users.}
    \label{fig:features}
\end{figure}

To demonstrate the effectiveness of GesIDNet in exploiting the data characteristics of gesture point clouds for gesture recognition and user identification, we visualize features extracted by GesIDNet with t-SNE~\cite{van2008visualizing}. 
As shown in Fig. \ref{fig:features} (a), for gesture recognition, low-level gesture features and high-level gesture features do not exhibit the same clear clusters for different gestures as the fusion features, indicating that the multilevel feature fusion further extracts effective unique features for different gestures by adaptively combining the low-level and high-level features.
Besides, as shown in Fig. \ref{fig:features} (b-1) \& (b-2), compared with gesture recognition, the low-level and high-level features extracted for user identification exhibit much lower clustering effect, indicating that features related to personalized motion patterns are more difficult to extract. 
However, after multilevel feature fusion, gesture samples belonging to the same user form into a clear cluster as demonstrated in Fig. \ref{fig:features} (b-3), indicating that GesIDNet achieves effective feature extraction for user identification.

\section{Implementation}
\label{sec:implementation}
\begin{figure}[t]
    \centering
    \includegraphics[width=0.49\textwidth]{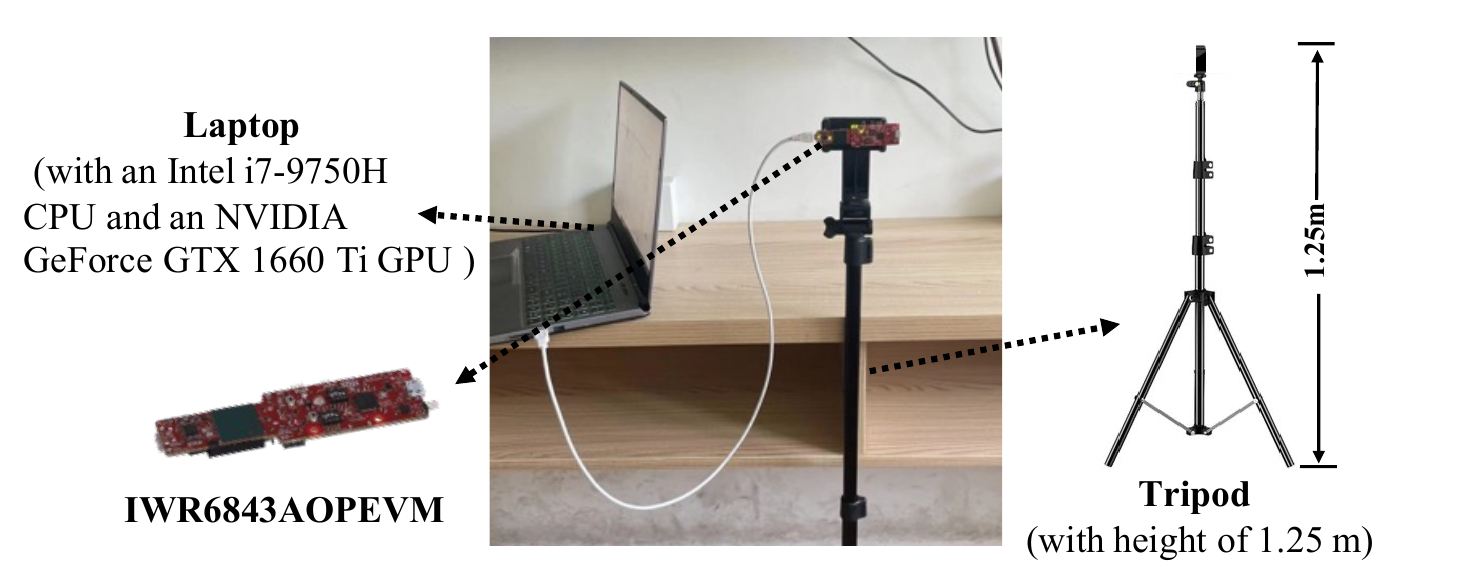}
    \caption{\textit{GesturePrint} utilizes an IWR6843AOPEVM mmWave radar sensor and a laptop for runtime inference. 
    The radar is positioned at a height of $1.25\,\text{m}$ with a tripod.
    }
    \label{fig:setup}
\end{figure}

\noindent\textbf{Hardware.} \textit{GesturePrint} consists of a commodity mmWave radar sensor, a laptop, and a back-end server. We use the IWR6843 antenna-on-package (AoP) evaluation module (EVM)~\cite{iwr6843} for data collection. The radar is mounted at a fixed height of $1.25\,\text{m}$ with a tripod. As shown in Fig. \ref{fig:setup}, the collected data is transmitted to the laptop with an Intel i7-9750H CPU and an NVIDIA GeForce GTX 1660 Ti GPU for signal processing and inference. Note that we can only use the CPU of the laptop for the inference phase~(see \S\ref{sec:overhead} for more details). The back-end server has an AMD EPYC 7402 24-Core CPU and an NVIDIA GeForce RTX 3090 GPU. The back-end server is used for model training and testing.
Besides, we also run the inference phase on Jetson Nano~\cite{nano} to evaluate the performance of \textit{GesturePrint} on the edge device.

\noindent\textbf{System Parameter Settings.} In \textit{GesturePrint} implementation, the parameter settings are as follows. The mmWave radar sensor operates in the $60\text{-}64\,\text{GHz}$ RF band and enables 3 TX and 4 RX antennae. Its frame rate is $10\,\text{fps}$, range resolution is $0.04\,\text{m}$, maximum unambiguous range is $8.2\,\text{m}$, maximum radial Doppler velocity is $2.7\,\text{m/s}$, and radial velocity resolution is $0.34\,\text{m/s}$. As for the parameters in the gesture segmentation module, the point number threshold window length $N$ is 50, the sliding motion detection window length $n$ is 10, and the minimum frame threshold $F_{Thr}$ is 8.
For noise canceling, the maximum distance $D_{max}$ is $1\,\text{m}$ and the minimum point number $N_{min}$ is 4.

\noindent\textbf{Model Training and Testing.} GesIDNet is implemented by using Python and PyTorch~\cite{paszke2019pytorch}. Typically, for all the datasets evaluated in the experiments, the split ratio of the training set and the test set is usually 8:2 with 5-fold cross-validation for reliable results. We employ the same split ratio and settings adopted by existing state-of-the-art methods~\cite{liu2022mtranssee,liu2020real,salami2022tesla,palipana2021pantomime} on the three public datasets used in the experiments to do a fair comparison with them on the gesture recognition task.

\section{Performance Evaluation}
\label{sec:evaluation}
\subsection{Datasets and Methodology}
\begin{table}[t]
\centering
\caption{Dataset summary.}
\resizebox{0.9\linewidth}{!}{
\begin{tabular}{cccc}
\toprule[2pt]
\multirow{1}{*}{\textbf{Dataset}} & \multirow{1}{*}{\textbf{Main Scenario}} & \multirow{1}{*}{\textbf{Gesture}} & \multicolumn{1}{c}{\textbf{User}}  \\ 
\midrule[2pt]
\multirow{2}{*}{Pantomime~\cite{palipana2021pantomime}} & \multicolumn{1}{c}{Office} & \multirow{2}{*}{21~(self-defined)} & \multicolumn{1}{c}{26}  \\ 
 & \multicolumn{1}{c}{Open} &  & \multicolumn{1}{c}{14} \\
\midrule

mHomeGes~\cite{liu2020real} & \multicolumn{1}{c}{Home} & \multicolumn{1}{c}{10~(self-defined)} & \multicolumn{1}{c}{8-14} \\ \midrule

mTransSee~\cite{liu2022mtranssee} & \multicolumn{1}{c}{Home} & \multicolumn{1}{c}{5~(self-defined)} & \multicolumn{1}{c}{32}   \\ \midrule
Self-collected & \multicolumn{1}{c}{Office} & \multirow{2}{*}{15~(ASL gestures)} & \multirow{2}{*}{17}  \\ 
(GesturePrint) & \multicolumn{1}{c}{Meeting Room} &  &  \\
 \bottomrule[2pt]
\end{tabular}
}
\label{table:dataset}
\end{table}

\subsubsection{Gesture Datasets} 
We evaluate \textit{GesturePrint} on four datasets that span diverse scenarios~(office, meeting room, home, and open space), user scales, and predefined gestures.
These datasets include our self-collected dataset~(GesturePrint), and three public datasets, i.e., the Pantomime dataset~\cite{palipana2021pantomime}, the mHomeGes dataset~\cite{liu2020real}, and the mTransSee dataset~\cite{liu2022mtranssee}.
All the public datasets contain self-defined gestures. In contrast, gestures in our self-collected dataset are standard gestures in ASL~(American Sign Language). 
Tab. \ref{table:dataset} provides a summary of these datasets.

    \noindent\textbf{Pantomime dataset~\cite{palipana2021pantomime}.} 
    It contains 22,291 samples of 21 self-defined gestures, including 9 easy single-arm gestures and 12 bimanual complex gestures, from 4-26 volunteers in different environments, positions, and articulation speeds. 
    
    \noindent\textbf{mHomeGes dataset~\cite{liu2020real}.} 
    It contains 22,000 samples from 8-14 participants performing 10 self-defined large arm movements at different anchor points ranging from $1.2\,\text{m}$ to $3.0\,\text{m}$ with an equal interval of $0.15\,\text{m}$. 
    
    \noindent\textbf{mTransSee dataset~\cite{liu2022mtranssee}.} It contains data from 32 participants at 13 anchor positions ranging from $1.2\,\text{m}$ to $4.8\,\text{m}$, and the predefined gestures are 5 self-defined arm motions.

\begin{figure}[t]
    \centering
    \begin{subfigure}[b]{0.24\textwidth}
        \centering
        \includegraphics[width=\linewidth]{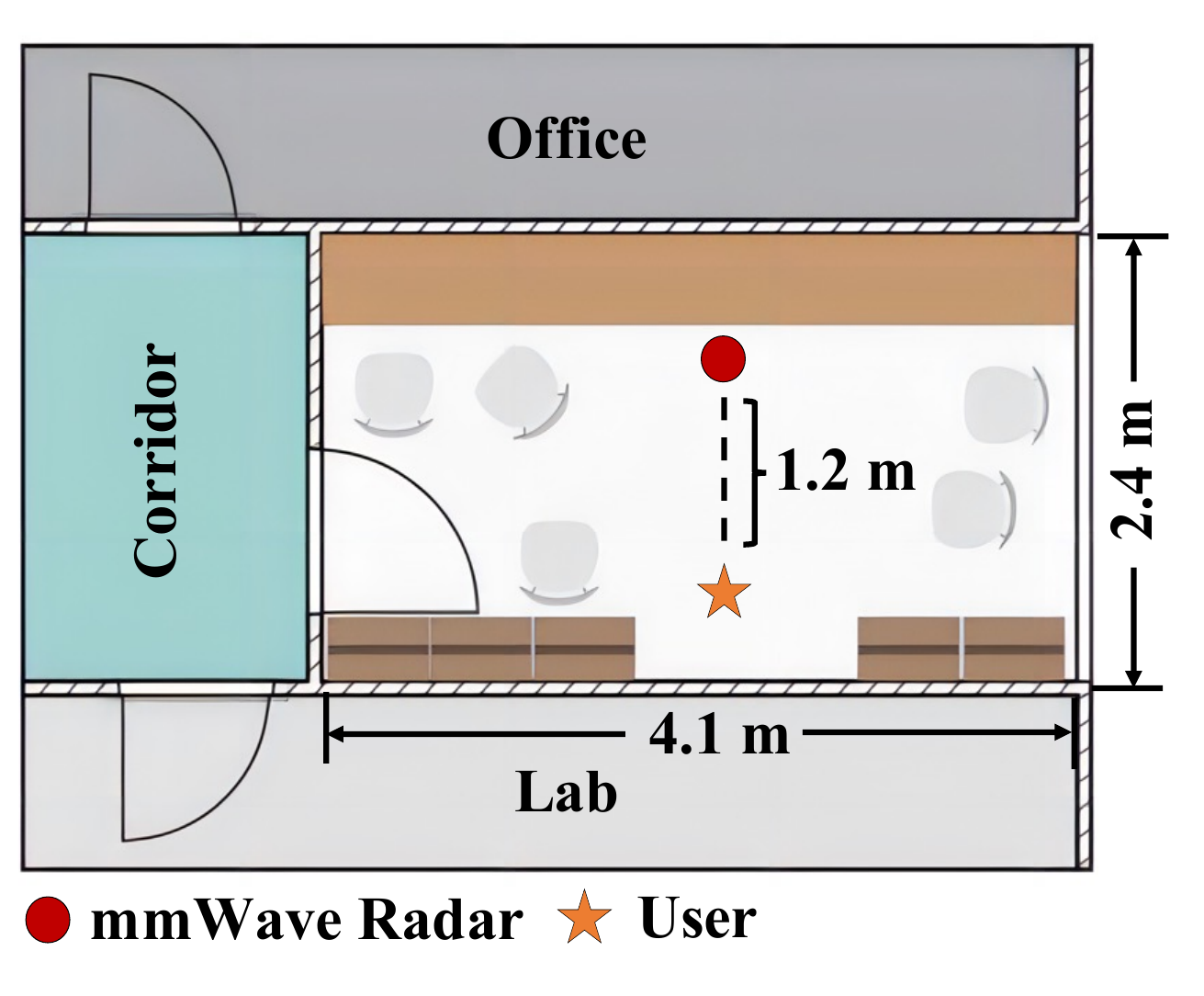}
        \caption{Office.}
        \label{fig:office}
    \end{subfigure}
    \begin{subfigure}[b]{0.24\textwidth}
        \centering
        \includegraphics[width=\linewidth]{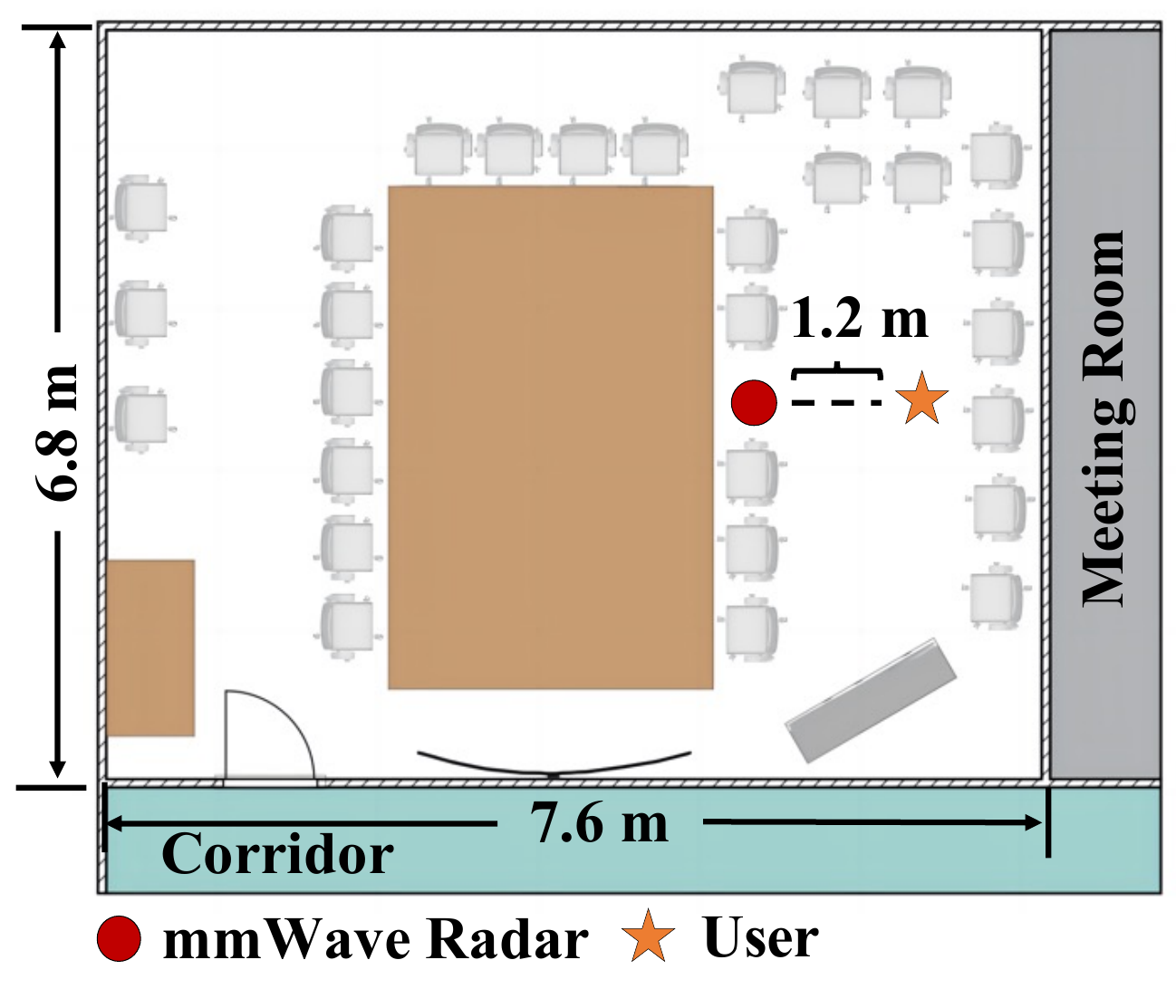}
        \caption{Meeting room.}
        \label{fig:meetingroom}
    \end{subfigure}
    \caption{
    Experimental environments.
    }
    \label{fig:env}
\end{figure}

\begin{figure}[t]
    \centering
    \includegraphics[width=0.485\textwidth]{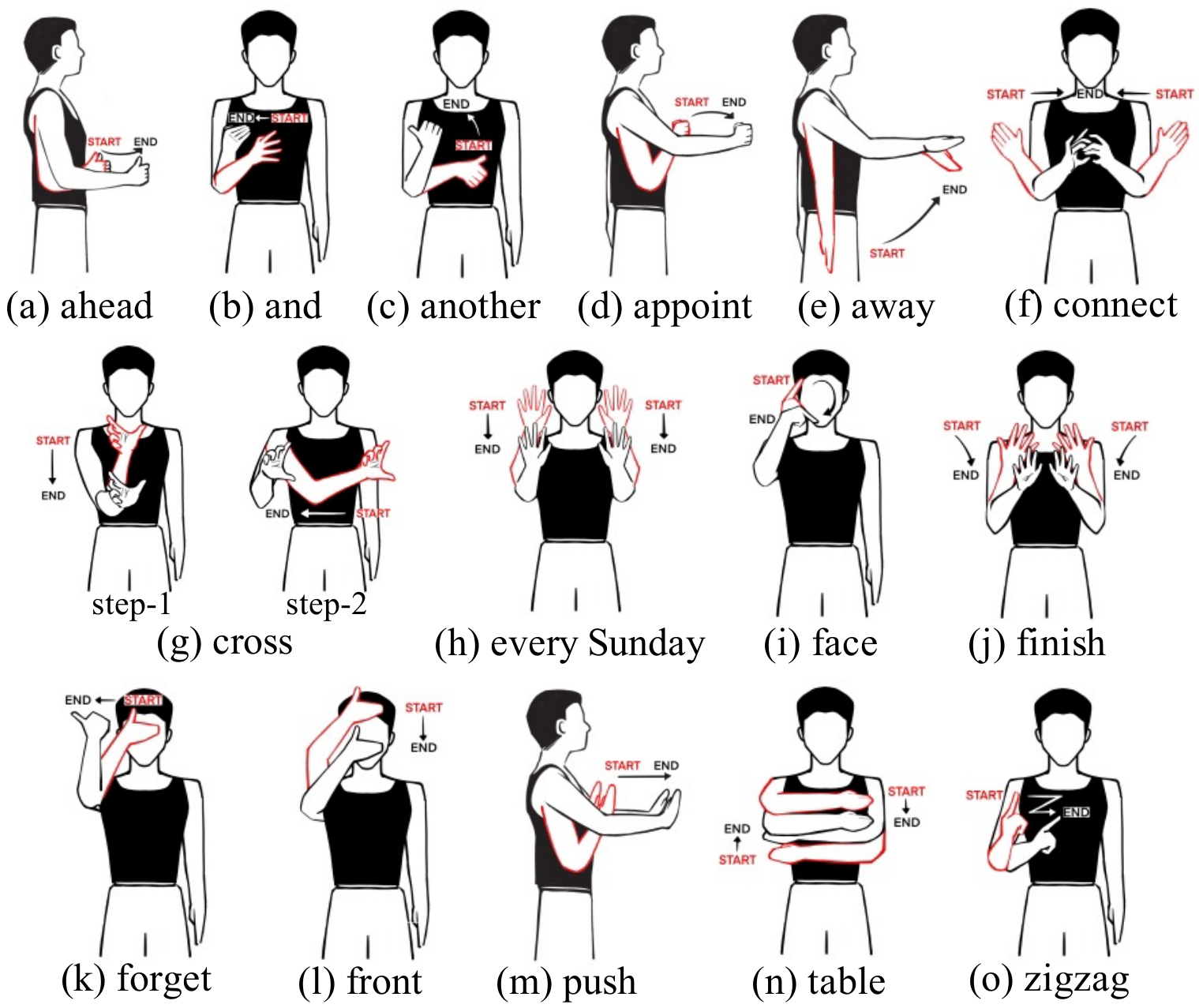}
    \caption{15 ASL signs in the GesturePrint dataset.
    The start of the gesture is colored in red; the end is colored in black.}
    \label{fig:gesture}
\end{figure}

    \noindent\textbf{Self-collected dataset (GesturePrint).} We conducted our data collection in two major environments, i.e., a small office and a large meeting room, as shown in Fig.~\ref{fig:env}. 
    As demonstrated in Fig.~\ref{fig:gesture}, we choose 15 basic and representative ASL signs from ASLLVD~\cite{athitsos2008american} as predefined gestures.
    The selected ASL signs contain 9 single-arm gesture motions and 6 bimanual gesture motions, including `ahead', `and', `another', `appoint', `away', `connect', `cross', `every Sunday', `face', `finish', `forget', `front', `push', `table' and `zigzag'.
    The chosen signs encompass a variety of combinations involving motions of the hands, forearms, elbows, and arms. We recruited 17 participants (10 females and 7 males) aged $20\text{–}27$, weight $40\text{–}85\,\text{kg}$, and height $1.55\text{–}1.80\,\text{m}$ to perform each gesture 12-25 times in each environment. Before collecting gesture data from the participants, we demonstrated how to perform the ASL gestures by playing a video illustrating standard gesture motion instructions, and the participants learned how to perform gestures based on their understanding. 
    During data collection, the participants were instructed to face the device at a distance of $1.2\,\text{m}$ and performed gestures continuously with a time interval of about 2-4 seconds between every two gesture motions.
    The data collection process spanned 16 days. For the two different environments, individual participants were involved in the data collection on different days.
    Finally, with 2 environments, 17 participants, and 15 ASL gestures, we collected 9,332 samples.

\subsubsection{Baseline}
We compare \textit{GesturePrint} with state-of-the-art methods, including the Pantomime network~(denoted as PanArch) from Pantomime \cite{palipana2021pantomime}, Tesla from Tesla-Rapture~\cite{salami2022tesla} (Pantomime's follow-up work that is evaluated on the Pantomime dataset), mGesNet from mHomeGes~\cite{liu2020real}, and mSeeNet from mTransSee \cite{liu2022mtranssee}.
All these existing gesture recognition studies are evaluated on their own gesture datasets as mentioned above. Differently, to evaluate the universality of \textit{GesturePrint}, we also apply it to the other three public mmWave-based point clouds gesture datasets apart from our self-collected dataset.
Moreover, we reuse these datasets from another dimension for user identification without any extra information.
Note that the state-of-the-art methods are not designed for the user identification task. Thus, we only compare \textit{GesturePrint} with them on the gesture recognition task.

\subsubsection{Evaluation Metrics}
We employ the following metrics to evaluate the performance of \textit{GesturePrint} on gesture recognition (GR) and user identification (UI).

\noindent\textbf{Accuracy (GRA \& UIA)}. GRA is the accuracy of the gesture recognition task, measuring the proportion of correctly predicted gesture samples among all samples. UIA is the accuracy of the user identification task. For our default identification mode, i.e., serialized mode, UIA is the average identification accuracy of all gestures. For the parallel mode, UIA is directly computed once on all gestures. The settings for other metrics related to user identification are similar.

\noindent \textbf{F1-Score (GRF1 \& UIF1)}. GRF1 is the F1-Score of recognizing gestures, considering both false positives and false negatives for each gesture.
UIF1 is the F1-Score of identifying users, taking into account both false positives and false negatives for each user.

\noindent \textbf{AUC (GRAUC \& UIAUC)}. GRAUC is the area under the receiver operating characteristic~(ROC) curve, and it measures the system's discriminatory power in recognizing gestures. UIAUC is employed for the user identification task.

\noindent \textbf{Equal Error Rate~(EER)}. It is the rate at which false positive rate~(FPR) equals false negative rate~(FNR).
The metric is used to measure user identification performance.
FPR is the probability that the system fallaciously identifies others as the target user; FNR is the probability that the system identifies the target user as others.

\begin{table*}[ht!]
\centering
\setlength{\tabcolsep}{1pt}
\caption{Overall gesture recognition and user identification performance.
GP represents \textit{GesturePrint}.
GP-S denotes GP-Serialized~(the default identification mode), while GP-P denotes GP-Parallel. SOTA represents state-of-the-art results.
\textit{GesturePrint} achieves comparable performance with SOTA on gesture recognition and enables effective user identification.}

\resizebox{1\linewidth}{!}{
\begin{tabular}{@{}ccccccccccccccccccc@{}}
\toprule[2pt]
\multirow{1}{*}{{Dataset}} & \multicolumn{6}{c}{{Self-collected (GesturePrint)}} & \multicolumn{6}{c}{{Pantomime}} & \multicolumn{3}{c}{{mHomeGes}}& \multicolumn{3}{c}{{mTransSee}}\\ 
\cmidrule(lr){2-7}\cmidrule(l){8-13} \cmidrule(l){14-16} \cmidrule(l){17-19}
\multirow{1}{*}{{Scenario}} & \multicolumn{3}{c}{{Office}} &  \multicolumn{3}{c}{{Meeting Room}}  & \multicolumn{3}{c}{{Office}} & \multicolumn{3}{c}{{Open}} & \multicolumn{3}{c}{{Home}} & \multicolumn{3}{c}{{Home}}\\
\cmidrule(lr){2-4}\cmidrule(lr){5-7}\cmidrule(l){8-10}\cmidrule(l){11-13} \cmidrule(l){14-16} \cmidrule(l){17-19}
\multirow{1}{*}{{Metrics}} & GRA & GRF1 & GRAUC& GRA & GRF1 & GRAUC& GRA & GRF1 & GRAUC& GRA & GRF1 & GRAUC& GRA & GRF1 & GRAUC& GRA & GRF1 & GRAUC\\
\midrule
 SOTA & \multicolumn{3}{c}{{/}}& \multicolumn{3}{c}{{/}} & 0.9714 & - & 0.9994~\cite{salami2022tesla} & 0.9612 & - & {0.9994}~\cite{palipana2021pantomime} & 0.9800~\cite{liu2020real} & - & - & 0.9800~\cite{liu2022mtranssee} & - & - \\
\multirow{1}{*}{GP (ours) }& {0.9822} &	{0.9821} & {0.9908} & {0.9887} & {0.9885} & {0.9942} & {0.9854} & {0.9846} & {0.9997} & {0.9662} & {0.9633} & 0.9993 & {0.9960} & {0.9957} & {0.9966} & {0.9988} & {0.9988} & {0.9992}\\
\midrule
\multirow{1}{*}{{Metrics}} & UIA & UIF1 & UIAUC & UIA & UIF1 & UIAUC & UIA & UIF1 & UIAUC & UIA & UIF1 & UIAUC & UIA & UIF1 & UIAUC & UIA & UIF1 & UIAUC \\
\midrule

 SOTA & \multicolumn{3}{c}{{/}}& \multicolumn{3}{c}{{/}}& \multicolumn{3}{c}{{/}}& \multicolumn{3}{c}{{/}}& \multicolumn{3}{c}{{/}}& \multicolumn{3}{c}{{/}}\\
\rowcolor{mygray} 
GP-S (ours) & \textbf{0.9926} & \textbf{0.9901} & \textbf{0.9947} & \textbf{0.9978} & \textbf{0.9972} & \textbf{0.9990} & \textbf{0.9985} & \textbf{0.9972} & \textbf{0.9987} & \textbf{0.9931} & \textbf{0.9902} & \textbf{0.9962} & \textbf{0.9933} & \textbf{0.9925} & \textbf{0.9969} & \textbf{0.9760} & \textbf{0.9707} & \textbf{0.9913} \\
GP-P (ours) & 0.9863 & 0.9811 & 0.9908 & 0.9909 & 0.9906 & 0.9954 & 0.9909 & 0.9908 & 0.9959 & 0.9865 & 0.9850 & 0.9928 & 0.9897 & 0.9895 & 0.9938 & 0.9398 & 0.9386 & 0.9742  \\
\bottomrule[2pt]
\end{tabular}

}
\label{table:gescls}
\end{table*}

\subsection{Experimental Results}
\subsubsection{\textbf{Overall Performance}} 
We evaluate the performance of \textit{GesturePrint} on four datasets, including three public datasets and our self-collected dataset.
To fairly compare \textit{GesturePrint} with state-of-the-art methods on the three public datasets, the distance between the radar and the user should be the same across all the datasets. 
Thus, we utilize gesture data collected at $1.2\,\text{m}$ in the mHomeGes, mTransSee, and self-collected datasets for overall performance comparison. 
However, the Pantomime dataset does not have such data. We utilize gesture data collected at $1\,\text{m}$ in the Pantomime dataset, which is the closest distance to $1.2\,\text{m}$. 
Besides, state-of-the-art methods on the Pantomime dataset~(PanArch and Tesla) are evaluated with the models trained on data from both the Office and Open subsets while being tested against different environments respectively. Thus, for gesture recognition, we follow their settings. 
For user identification, we train and test on the Office and Open subsets separately, since the participants in these two environments are different in the Pantomime dataset.

\begin{figure}[t]
    \centering
    \includegraphics[width=0.49\textwidth]{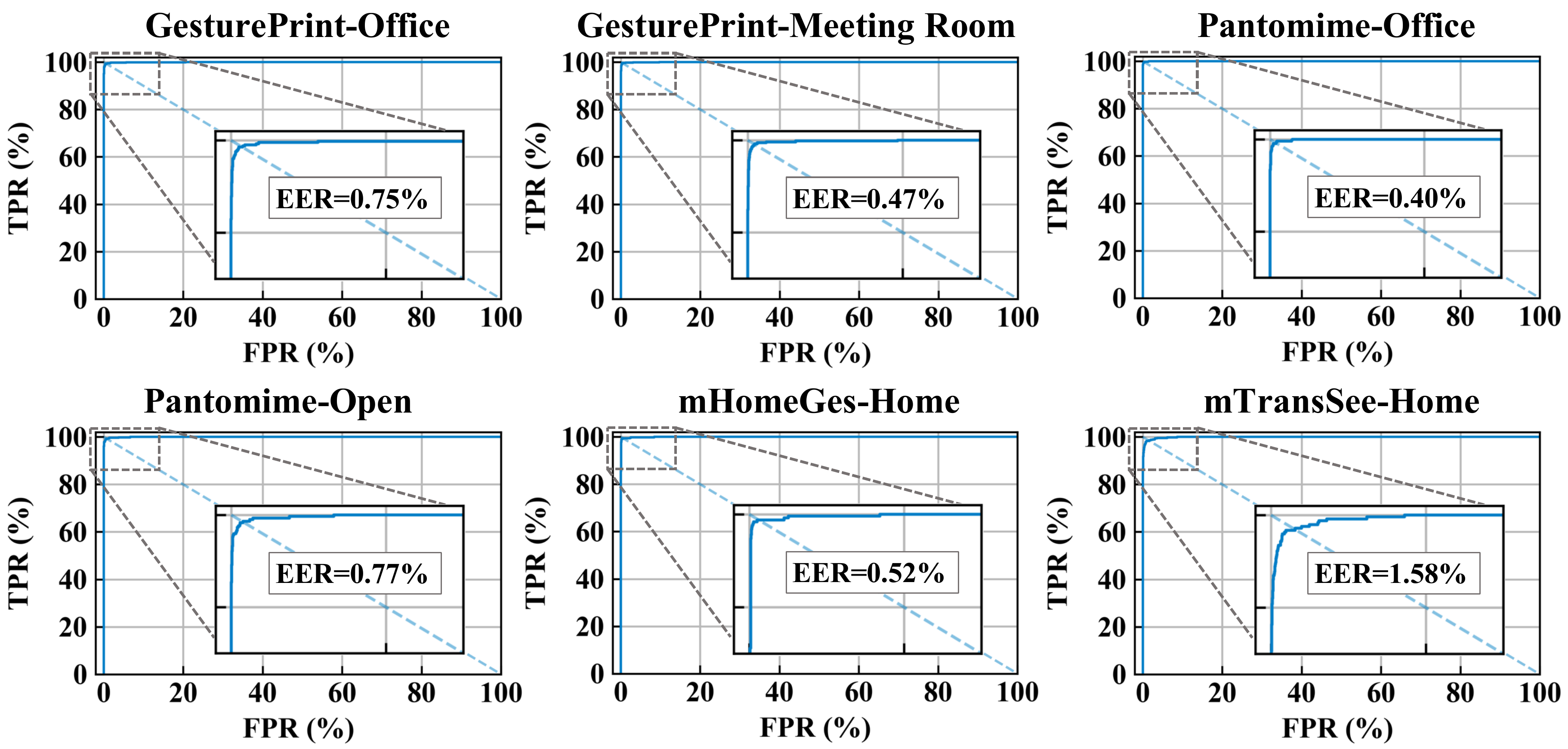}
    \caption{The ROC curves and EER results of user identification. \textit{GesturePrint} achieves an average result of 0.75\% EER.}
    \label{fig:eer}
\end{figure}
\noindent\textbf{Gesture Recognition.} To validate \textit{GesturePrint}'s ability to achieve reliable gesture recognition, we use gesture samples collected from all participants for training and testing. 
As shown in Tab.~\ref{table:gescls}, the overall recognition accuracy is above 96\% for gesture recognition across all the datasets. 
On our self-collected dataset, \textit{GesturePrint} achieves GRA of 98.22\% and 98.87\% in the office and meeting room environments, respectively.
Furthermore, we evaluate \textit{GesturePrint} on the other three datasets. 
To be specific, on the Pantomime dataset, we achieve 98.54\% GRA in the office environment, and 96.62\% GRA in the open space environment. On the mHomeGes dataset, we achieve 99.60\% GRA at the $1.2\,\text{m}$ anchor position. On the mTransSee dataset, we achieve 99.88\% GRA at $1.2\,\text{m}$.
Compared with the state-of-the-art results on the three datasets~\cite{palipana2021pantomime,salami2022tesla,liu2020real,liu2022mtranssee}, \textit{GesturePrint} achieves comparable results with 0.5\%-1.88\% accuracy improvements. 
Besides, the system maintains GRF1 above 0.96 and GRAUC exceeding 0.99 across all the datasets.
All the results indicate \textit{GesturePrint}'s effectiveness and universality on the gesture recognition task.

\noindent\textbf{User Identification.} As for our default identification mode, i.e., serialized mode, the overall identification performance is over 97\% as shown in Tab.~\ref{table:gescls}. On the self-collected dataset, \textit{GesturePrint} achieves 99.26\% UIA and 99.78\% UIA in the office and meeting room, respectively. 
Furthermore, we evaluate \textit{GesturePrint} on the other three datasets, and find it can maintain high accuracy for user identification based on different gestures. 
The results show that \textit{GesturePrint} is effective with different user scales. 
With the scale of users increasing, \textit{GesturePrint} can still successfully maintain 97.60\% UIA for recognizing 32 users on the mTransSee dataset.
For \textit{GesturePrint} with parallel mode, the UIA results have a slight drop within 4\% compared with the serialized mode, indicating that our method can indeed extract personal motion patterns even across different gestures.
Besides, the system consistently achieves reliable UIF1 and UIAUC across all the datasets. 
Moreover, as shown in Fig. \ref{fig:eer}, \textit{GesturePrint} achieves an average result of 0.75\% EER across all the datasets, with none exceeding 1.6\% EER.
The above results indicate the system's effectiveness in identifying users.

Overall, \textit{GesturePrint} achieves superior performance (GRA and GRF1 above 96\%, GRAUC above 99\%; UIA and UIF1 above 97\%, UIAUC above 99\%, EER below 1.6\%) across all the datasets, indicating that it is effective in achieving gesture recognition and enabling gesture-based user identification.

\begin{figure}[t]
    \centering
    \includegraphics[width=0.49\textwidth]{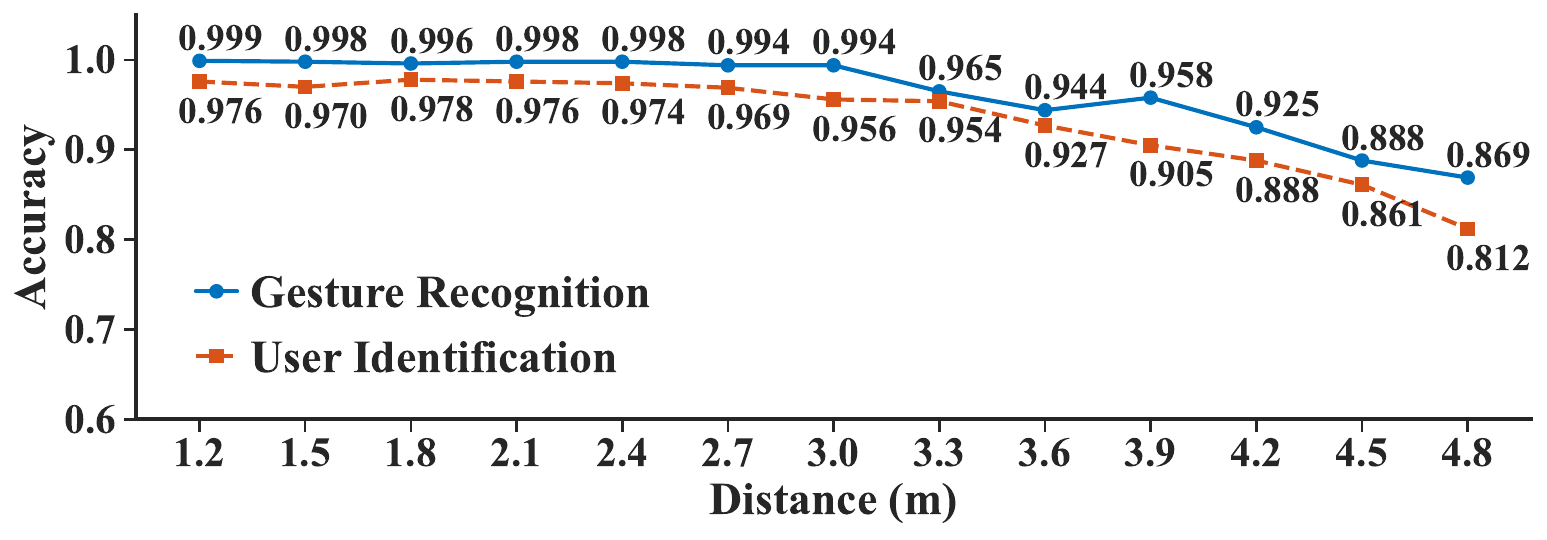}
    \caption{Impact of distance on \textit{GesturePrint}'s performance, the metrics in the figure are GRA and UIA. 
    }
    \label{fig:dis_ges}
\end{figure}

\subsubsection{\textbf{Impact of Distance}}
\label{sec:disimpact}
In this part, we first discuss the impact of the distance between the radar and the user on \textit{GesturePrint}'s performance.
Subsequently, we further evaluate \textit{GesturePrint} within a certain range of distance variations.

We evaluate \textit{GesturePrint} across all the anchor positions within the mTransSee dataset, and the results are demonstrated in Fig.~\ref{fig:dis_ges}.
As for the overall performance of \textit{GesturePrint} at different positions, \textit{GesturePrint} can maintain both reliable recognition performance ($\geq$ 94.4\% GRA) and identification performance ($\geq$ 92.7\% UIA) within a distance of $3.6\,\text{m}$. 
Besides, although when working at a distance exceeds $3.9\,\text{m}$, \textit{GesturePrint}'s performance is not as good as that within $3.6\,\text{m}$, the system can still achieve 86.9\% GRA and 81.2\% UIA at the furthest distance of $4.8\,\text{m}$. 
The performance degradation of \textit{GesturePrint} working at a distant position is because the point number captured by the radar rapidly decreases with increasing distance due to the limitation of the hardware settings~\cite{liu2020real}. 

\begin{figure}[t]
    \centering
    \includegraphics[width=0.49\textwidth]{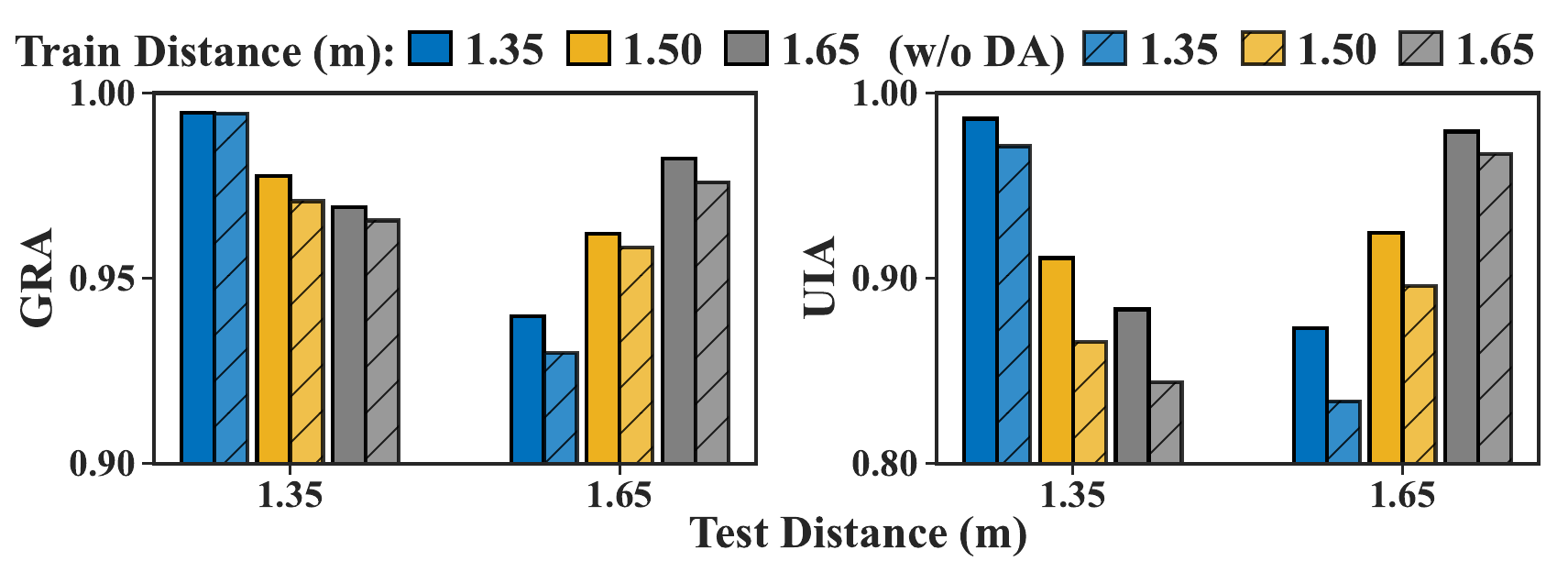}
    \caption{Distance robustness experiments on different anchor positions~({$1.35\,\text{m}$, $1.5\,\text{m}$, $1.65\,\text{m}$}). DA: data augmentation.}
    \label{fig:crossdis}
\end{figure}

We further evaluate the distance robustness of \textit{GesturePrint} on a subset comprising three different anchor positions~($1.35\,\text{m}$, $1.5\,\text{m}$, and $1.65\,\text{m}$) of the mHomeGes dataset. 
We use data from different distances as the training set and the test set, respectively.
As shown in Fig.~\ref{fig:crossdis}, the results indicate that \textit{GesturePrint} can achieve reliable performance on both gesture recognition and user identification when confronted with positions that are previously unseen during training.

Considering the limitation of the device configuration, distance robustness of \textit{GesturePrint}, and practical application scenarios, when users try to interact with \textit{GesturePrint} from a distant position, \textit{GesturePrint} can remind the user to step closer and enter the area where it can work reliably.

\begin{figure}[t]
    \centering
    \includegraphics[width=0.485\textwidth]{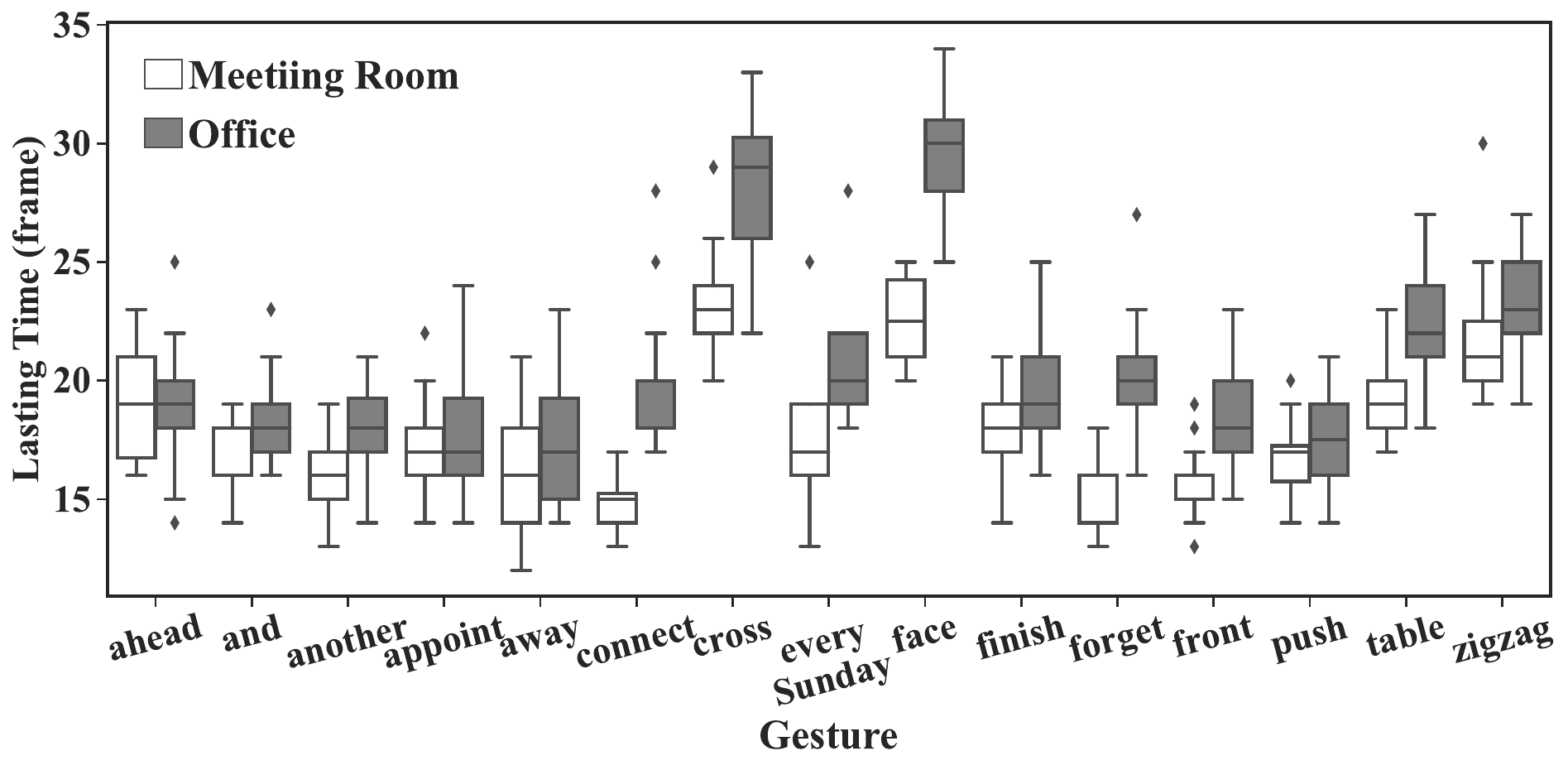}
    \caption{Lasting time of gesture motions performed by the same user. Users exhibit variations in their motion speed when repeatedly executing the same gesture.
    }
    \label{fig:speed}
\end{figure}

\subsubsection{\textbf{Impact of Motion Speed}} An implicit insight behind \textit{GesturePrint} is that there is some information in the gesture point clouds that can reflect one's specific motion style and behavior personality, thus we can design effective network architecture to extract high-dimension features containing such information for user identification. 
However, there is a question of whether such features will distort when users perform the same gesture at different speeds, leading to the extracted features becoming ineffective. 
As shown in Fig.~\ref{fig:speed}, it is inevitable that users slightly change their motion speed unconsciously, evidenced by the gesture's lasting time measured by the frame number.
Based on the observation, we further evaluate \textit{GesturePrint} on a subset of the Pantomime dataset, which includes three different motion speeds.
The results demonstrate that, with deliberate change in the motion speed, \textit{GesturePrint} can still reach 97.73\% GRA and 98.81\% UIA. 
Thus, even if users change motion speeds when performing gestures, \textit{GesturePrint} can still extract effective features for gesture recognition and user identification, respectively.

\subsubsection{\textbf{Ablation Experiment}} 
\begin{figure}[t]
    \centering
    \begin{subfigure}[b]{0.49\textwidth}
        \centering
        \includegraphics[width=1.\linewidth]{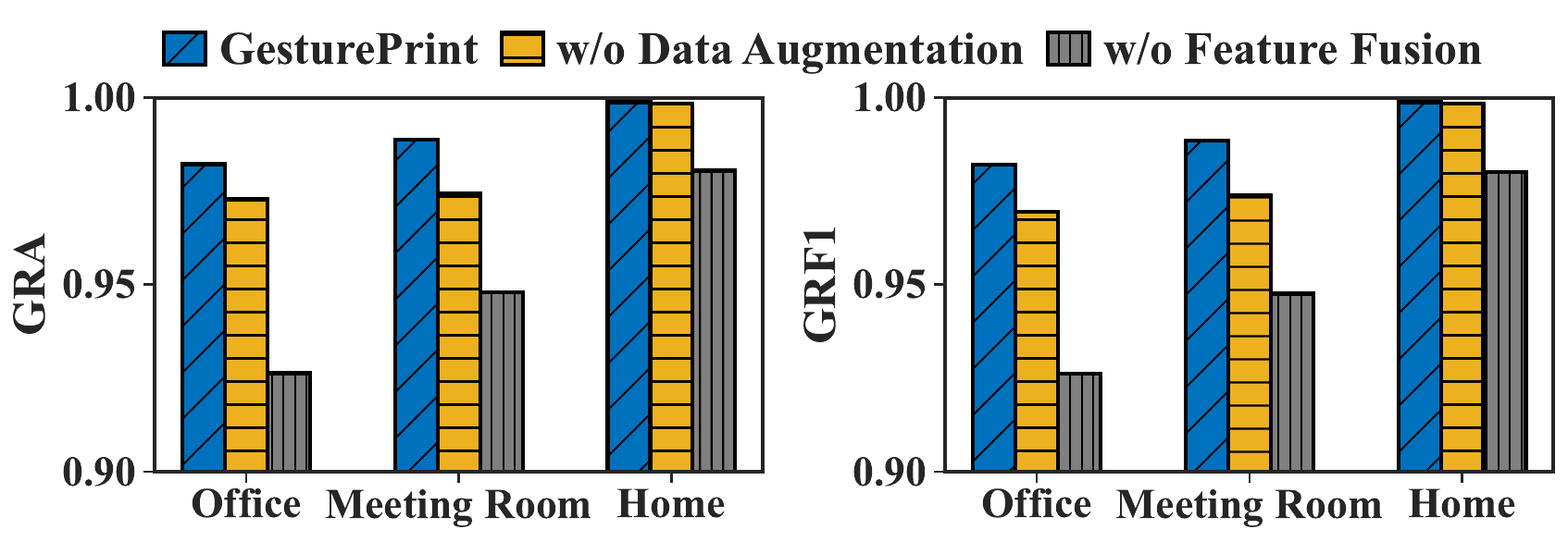}
        \caption{The gesture recognition performance.}
        \label{fig:ablation_gesture}
    \end{subfigure}
    \hfill
    \begin{subfigure}[b]{0.49\textwidth}
        \centering
        \includegraphics[width=1.\linewidth]{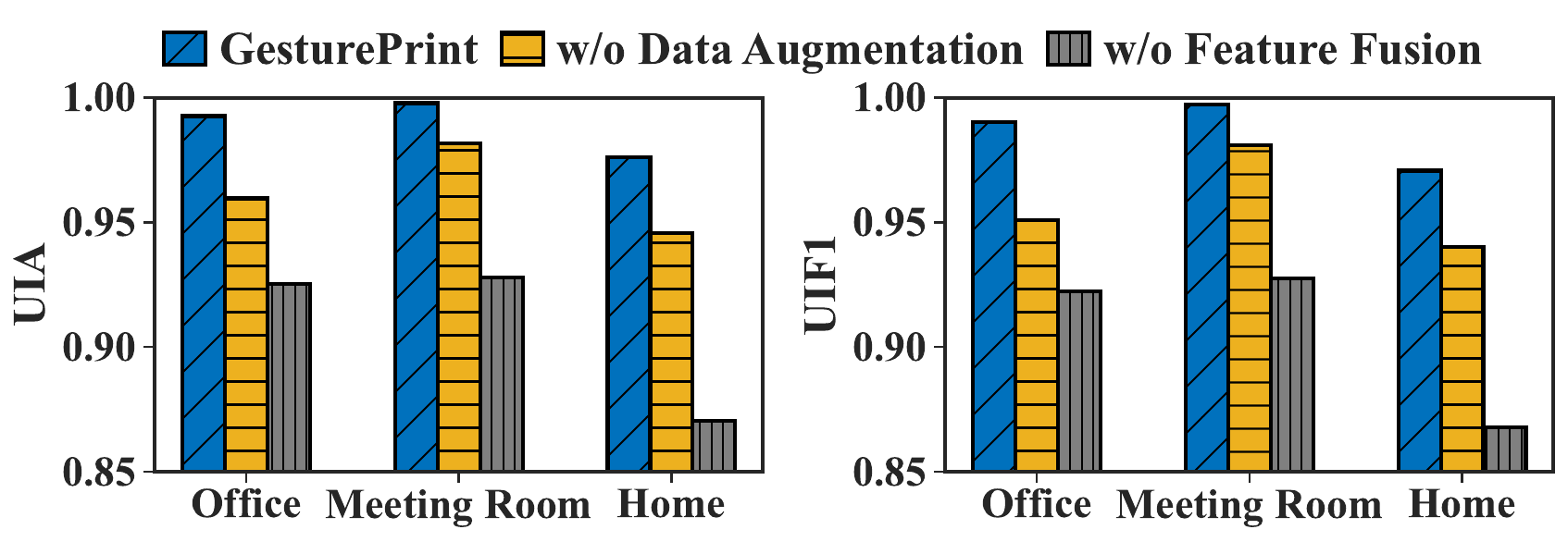}
        \caption{The user identification performance.}
        \label{fig:ablation_user}
    \end{subfigure}
    \caption{Ablation experiments about data augmentation and multilevel feature fusion. `Office' and `Meeting Room' are subsets from the GesturePrint dataset and `Home' is from the mTransSee dataset.
    The results show that these components effectively contribute to \textit{GesturePrint}'s performance.}
    
    \label{fig:ablation}
\end{figure}

In this part, we discuss the impacts of data augmentation in the data preprocessing stage and the attention-based multilevel feature fusion module in GesIDNet through ablation experiments.  
As shown in Fig.~\ref{fig:ablation}, data augmentation and the feature fusion module can improve the performance of \textit{GesturePrint} on both the gesture recognition and user identification tasks.
The designed feature fusion module contributes considerably to the performance improvement, and the contribution is extremely obvious in the scenarios with a large user scale, e.g., the `Home' scenario from the mTransSee dataset.
Moreover, data augmentation enhances \textit{GesturePrint}'s robustness to distance variations. 
As shown in Fig.~\ref{fig:crossdis}, without data augmentation, \textit{GesturePrint}'s performance decreases at distances unseen during training.

\subsubsection{\textbf{Time Consumption}}
\label{sec:overhead}
The time consumption of \textit{GesturePrint} to process a single gesture sample comprises the data preprocessing time and the classification inference time.
To evaluate \textit{GesturePrint}'s efficiency, we measure the average preprocessing time based on our self-collected dataset, and the classification inference time is measured by averaging the inference results over 500 runs.

The experimental results show that the average preprocessing time is $405.93\,\text{ms}$.
As for the classification inference time, when using the CPU of the laptop alone, the average inference time for recognition and identification is $677.14\,\text{ms}$.
By using the GPU of the laptop, the inference time can be reduced to $530.99\,\text{ms}$.
The average total time consumption for processing a single gesture sample is $936.92\,\text{ms}$~($0.94\,\text{s}$), while the average gesture duration in our dataset is $2.43\,\text{s}$. Thus, we believe that the time consumption of \textit{GesturePrint} can meet the requirements of most gesture interaction applications.

Besides, we deploy the models on Jetson Nano~\cite{nano} to evaluate the inference time on an edge device. 
The inference time is $1.58\,\text{s}$ in total for the gesture recognition and user identification tasks.
There are some other edge devices delivering superior computing power and AI performance compared with Jetson Nano.
For example, the newly launched Jetson Orin Nano Series~\cite{JetsonOrinNano} can offer up to 80$\times$ the performance of the standard Jetson Nano we utilized in the experiments.
Implementing \textit{GesturePrint} on these more powerful devices could further reduce the inference time.
This indicates the potential for applications of \textit{GesturePrint} on the edge.

\section{Discussion}
\label{sec:discussion}
\subsubsection{\textbf{Multi-person Scenarios}}

\begin{figure}[t]
    \centering
    \begin{subfigure}[b]{0.24\textwidth}
        \centering
        \includegraphics[width=\linewidth]{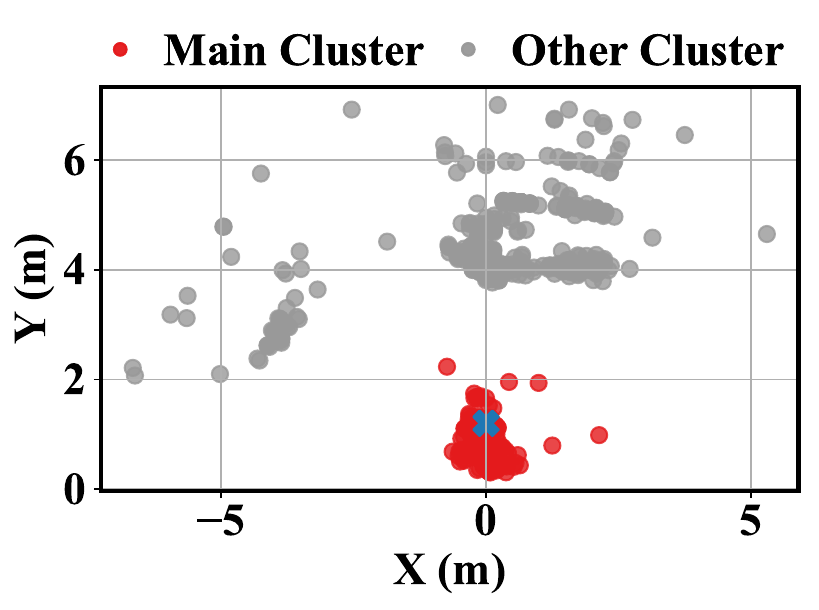}
        \caption{Someone else walks past behind the user.}
        \label{fig:multiperson_walk}
    \end{subfigure}
    \hfill
    \begin{subfigure}[b]{0.24\textwidth}
        \centering
        \includegraphics[width=\linewidth]{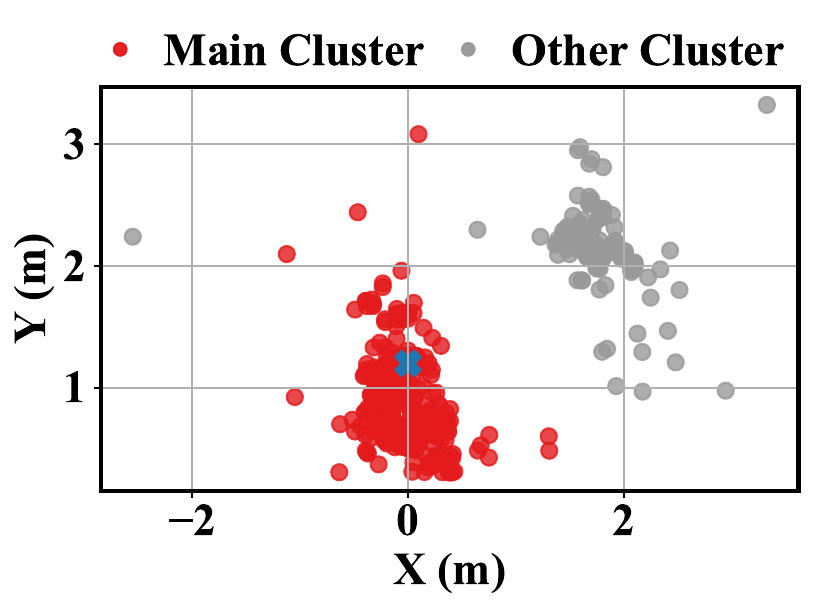}
        \caption{Someone else performs gestures at the same time.}
        \label{fig:multiperson_static}
    \end{subfigure}
    \caption{
    \textit{GesturePrint} separates main point clusters related to the target user from other clusters. The blue cross in the figures denotes the position of the user.}
    \label{fig:multiperson}
\end{figure}

Multiple persons may be active in the scenario when a user performs gestures, which affects the system's performance adversely.
Although \textit{GesturePrint} is mainly designed to unleash the potential of existing gesture recognition systems by enabling user identification, it can handle some common multi-person cases.
By distinguishing main point clusters and noise canceling, \textit{GesturePrint} reduces distractions from others when the user interacts with the system.
Fig.~\ref{fig:multiperson} shows two common cases, i.e., someone else walking around or performing gestures.
\textit{GesturePrint} can figure out the point clusters related to the user during the data preprocessing stage.
Note that the minimum distinguishable distance between the user and others is primarily determined by the parameters of DBScan, which can be measured and configured accordingly before the deployment of \textit{GesturePrint}.
Furthermore, in practice, a work zone can be predefined to remind users to perform gestures within a specific area for reliable identification as suggested in \S\ref{sec:disimpact}. 
In this way, the impact of other people outside the area can be further mitigated.
As for those who perform gestures very close to the user on purpose, the user could easily notice such interference.

In future work, \textit{GesturePrint} could be extended to address more complex multi-user scenarios, where the system needs to simultaneously process interaction gestures from multiple users.
The multi-user detection method proposed by $m^{3}Track$~\cite{kong2022m3track} provides a solution to improve \textit{GesturePrint}.

\subsubsection{\textbf{Cross-domain Scenarios}}
Cross-domain adaptability is a pervasive and important issue in sensing.
As discussed in RF-Net~\cite{ding2020rf}, RF sensing is more sensitive to environmental changes as it detects both the subject and background.
Consequently, exploring the impacts of different positions and distinct environments on our system is beneficial. 
We have conducted experiments under cross-distance and cross-environment settings to gain insights.

As discussed in \S\ref{sec:disimpact}, Fig.~\ref{fig:crossdis} demonstrates that \textit{GesturePrint} maintains over 93\% GRA and 87\% UIA under cross-distance settings, indicating the system's capability and potential for handling gesture interactions in unseen positions.
For conducting cross-environment experiments, we utilize the two distinct environments in our self-collected dataset, training models on data from Office/Meeting Room while testing them on Meeting Room/Office.
The results indicate that \textit{GesturePrint} maintains over 90\% GRA and about 75\% UIA under two cross-environment situations.
The performance decline resulting from cross-environment challenges can be mitigated by fine-tuning the models with data collected from the target environment. 

Currently, \textit{GesturePrint} is mainly designed for scenarios where authorized users interact with devices, such as smart homes/buildings. In these scenarios, mmWave radar sensors are typically placed in a fixed position, which allows for a predefined work zone.
Future work that addresses the cross-domain issue could extend \textit{GesturePrint} to address broader application scenarios.

\section{Conclusion}
\label{sec:conclusion}
This paper proposes \textit{GesturePrint}, a one-stop solution including gesture recognition and further gesture-based user identification based on the mmWave radar.
\textit{GesturePrint} extracts unique gesture features for gesture recognition and personalized motion pattern features for user identification, respectively. 
We design a signal processing stage to efficiently preprocess the raw data to obtain gesture point clouds.
Furthermore, we design a network architecture GesIDNet with an attention-based multilevel feature fusion mechanism to extract effective features from gesture point clouds. 
Extensive experiments demonstrate \textit{GesturePrint}'s effectiveness in achieving reliable gesture recognition and enabling user identification.
To summarize, \textit{GesturePrint} is an effective system that unleashes the potential of mmWave-based gesture recognition systems with user identification at a minor cost.
\section*{Acknowledgment}
Chaojie Gu is the corresponding author.
This work was completed when Keyi Wang was a final-year undergrad at Zhejiang University.
Thank the anonymous reviewers for providing valuable feedback on this work.
Thank the volunteers for participating in the data collection.
This research is supported by the National Natural Science Foundation of China under Grants No. (62302439, U23A20296), the Key Research and Development Program of Zhejiang Province (2021C03098), and the Fundamental Research Funds for the Central Universities (226-2023-00111, 226-2024-00004).

\balance
\bibliographystyle{IEEEtran}
\bibliography{ref}

\end{document}